\documentclass[11pt, a4paper]{article}
\pdfoutput=1

\usepackage{amsmath,amssymb}
\usepackage{comment}
\usepackage{multirow}
\usepackage[utf8]{inputenc}
\usepackage{bm}
\usepackage{cite}
\usepackage[footnotesize]{caption}

\usepackage{ifpdf}
\ifpdf        
  \usepackage{graphicx, hyperref, xcolor}     
\else     
  \usepackage[dvipdfmx]{graphicx, hyperref, xcolor}     
 \fi
\usepackage{subcaption}

\usepackage{tikz}
\usepackage{tikz-feynman}
\tikzfeynmanset{compat=1.0.0}
 
\definecolor{rossoferrari}{HTML}{D9073D}
\definecolor{mediumblue}{HTML}{0000CD}
\definecolor{forestgreen}{HTML}{228B22}
\definecolor{desy_blue}{HTML}{009EE2}
\definecolor{desy_orange}{HTML}{FD8800}
\hypersetup{
setpagesize=false,
bookmarksnumbered=true,%
bookmarksopen=true,%
colorlinks=true,%
linkcolor=rossoferrari,
urlcolor=mediumblue,
citecolor=mediumblue,
linktocpage=false,
}

\usepackage[height=8.85in,width=6.4in]{geometry}

\usepackage{fourier}

\makeatletter
\@addtoreset{equation}{section}

\makeatother

\newcommand{\abs}[1]{\left\lvert #1 \right\rvert}
\newcommand{\dd}{\mathrm{d}}

\begin{document}

\begin{titlepage}

\begin{center}

\hfill DESY 20-031

\vskip 1.in

{\Huge \bf
Higgs Inflation as Nonlinear Sigma Model \\
and Scalaron as its $\sigma$-meson\\
}

\vskip .8in

{\Large
Yohei Ema, Kyohei Mukaida, Jorinde van de Vis
}

\vskip .3in

{\em DESY, Notkestra{\ss}e 85, D-22607 Hamburg, Germany}

\end{center}
\vskip .6in

\begin{abstract}
\noindent
We point out that a model with scalar fields with a large nonminimal coupling to the Ricci scalar, such as Higgs inflation, can be regarded as a nonlinear sigma model (NLSM). 
With the inclusion of not only the scalar fields but also the conformal mode
of the metric, our definition of the target space of the NLSM is invariant under the frame transformation.
We show that the $\sigma$-meson that linearizes this NLSM 
to be a linear sigma model (LSM) corresponds to the scalaron, 
the degree of freedom associated to the $R^2$ term in the Jordan frame.
We demonstrate that quantum corrections inevitably induce this $\sigma$-meson in the large-$N$ limit, 
thus providing a frame independent picture for the emergence of the scalaron. 
The resultant LSM only involves renormalizable interactions and hence its perturbative unitarity holds up to the Planck scale unless it hits a Landau pole, which is in agreement with the renormalizability of quadratic gravity. 

\end{abstract}

\end{titlepage}

\renewcommand{\thepage}{\arabic{page}}
\setcounter{page}{1}

\tableofcontents
\pagebreak
\renewcommand{\thepage}{\arabic{page}}
\renewcommand{\thefootnote}{$\natural$\arabic{footnote}}
\setcounter{footnote}{0}

\section{Introduction}
\label{sec:intro}

Cosmic inflation is a successful paradigm for the description of the very early Universe.
While solving the flatness and horizon problems in Big Bang cosmology, its accelerated expansion of the Universe provides an origin of anisotropies in the cosmic microwave background (CMB) and gives rise to primordial gravitational waves.
Such quasi-de Sitter phase is realized once we have a scalar field, the so-called inflaton, slowly rolling down its potential during inflation.
The combined bounds from the current observations in the $(n_{s},r)$-plane~\cite{Akrami:2018odb} imply a concave potential for the inflaton.

Inflation caused by the Standard Model Higgs stands out as an attractive candidate model among many others because of its minimality.
To have successful inflation, a nonminimal coupling to gravity is introduced in Refs.~\cite{Futamase:1987ua,CervantesCota:1995tz,Bezrukov:2007ep}:
\begin{align}
	\mathcal{L}_{\xi} = \xi R \abs{H}^{2},
	\label{eq:defining}
\end{align}
where $H$ is the Standard Model Higgs doublet and $\xi$ is the nonminimal coupling of $H$ to the Ricci scalar~$R$.
This term modifies the Higgs quartic potential for a large field value of the Higgs $\abs{H} \gtrsim M_{P} / \xi$,  
in perfect agreement with the aforementioned observational bound~\cite{Akrami:2018odb}.
To produce a curvature perturbation of the right magnitude, the Higgs quartic coupling $\lambda$ and the nonminimal coupling $\xi$ should fulfill $\xi^{2} \simeq 2 \times 10^{9} \lambda$, implying $\xi \gg 1$ unless $\lambda$ is extremely small.\footnote{
	See Refs.~\cite{Hamada:2014iga,Bezrukov:2014bra} for critical Higgs inflation that
	has a tiny $\lambda$ via the running.
}

Classically, such a large value of $\xi$ is just a choice of a parameter. However, quantum corrections
induce other operators associated with this large coupling via a renormalization group (RG) flow. 
In particular, by computing scalar one-loop diagrams in the Jordan frame, 
one finds an enhancement of the $R^2$
term for $\xi \gg 1$~\cite{Donoghue:1995cz,Salvio:2015kka,Calmet:2016fsr,Ema:2017rqn,Ghilencea:2018rqg,Ema:2019fdd}.
\begin{align}
	\mathcal{L}_{\alpha} = \alpha R^{2}, \qquad \frac{\dd \alpha}{\dd \ln \mu} = - \frac{N}{1152 \pi^{2}} \left( 6 \xi + 1 \right)^{2},
	\label{eq:jordan-running}
\end{align}
where $N$ counts the number of real scalar fields, \emph{i.e.}, $N = 4$ for Higgs-inflation.
This $R^{2}$ term makes the scalar part of the metric dynamical, 
corresponding to the so-called \emph{scalaron}, whose mass is $m_{s}^{2} \sim M_{P}^{2} / \alpha$~\cite{Starobinsky:1980te,Barrow:1983rx,Whitt:1984pd,Barrow:1988xh}.
Although one may choose $\alpha$ to be small at a particular scale, this never holds for the entire range of energy scales
due to the RG running, implying its \emph{typical} value is $\alpha \sim \xi^{2} \gg 1$.\footnote{
	Strictly speaking, the running coupling at a scale $\mu$ involves a numerical factor and 
	a log term as $\alpha_{\mu} \sim 10^{-2} \xi^{2} \ln \Lambda/\mu$ (for $N=4$). 
	Here $\Lambda$ is the scale at which $\alpha$ vanishes 
	(analogous to $\Lambda_\mathrm{QCD}$ of QCD).  
	Throughout this paper, we omit this numerical factor in the estimation 
	because it is translated into at most an order one factor in the scalaron mass, 
	$m_{s} \sim M_{P}/\alpha^{1/2}$.
}
Note that the Renormalization Group Equation (RGE) of the other operator appearing at the same loop level, $\mathcal{L}_{\alpha_{2}} = \alpha_{2} (R_{\mu\nu} R^{\mu\nu} - R^2 / 3)$, does not depend on $\xi$.\footnote{
	The beta functions of $\alpha$ and $\alpha_2$ are computed in, \emph{e.g.}, 
	Refs.~\cite{Fradkin:1981iu,Elizalde:1993ee,Elizalde:1993ew,Salvio:2014soa,Salvio:2017qkx,Salvio:2018crh,
	Codello:2015mba,Myrzakulov:2016tsz,Markkanen:2018bfx,Kubo:2018kho,Ema:2019fdd},
	although the sign of the beta function of $\alpha$ is wrong in some references.
	Eq.~\eqref{eq:jordan-running} agrees with, \emph{e.g.}, 
Refs.~\cite{Fradkin:1981iu,Salvio:2014soa,Salvio:2017qkx,Salvio:2018crh,Markkanen:2018bfx,Kubo:2018kho,Ema:2019fdd}.
} Therefore, this operator is less important than the $R^2$ term
and can be neglected below the Planck scale in the limit $\xi \gg 1$.

Although physics should be independent under a frame transformation,
all the above observations related to the appearance of the light scalaron at $\xi \gg 1$ 
rely on the Jordan frame analysis.
In order to illustrate this point, let us move to, \textit{e.g.}, the Einstein frame.
In the Einstein frame, there is no large nonminimal coupling 
between the Higgs and the Ricci scalar.
It follows that there is no large enhancement of the $R^2$ term
as its RGE is now given by Eq.~\eqref{eq:jordan-running} with $\xi = 0$.
The large nonminimal coupling in the Jordan frame~\eqref{eq:defining} instead appears 
in the kinetic term of the Higgs in the Einstein frame:
\begin{align}
	\mathcal{L}_{\mathrm{kin}} = 
	\frac{1}{\left(1+ 2\xi \abs{H}^2/M_P^2\right)^2}
	\left[
	\left(1+ \frac{2\xi \abs{H}^2}{M_P^2}\right) \abs{\partial H}^2
	+ \frac{3 \xi^2}{M_P^2}\left(\partial \abs{H}^2 \right)^2
	\right].
	\label{eq:nlsm-intro}
\end{align}
Notice that, since the Higgs field contains four degrees of freedom, 
one cannot canonically normalize all the components at the same time.\footnote{
	One should not think that the kinetic term can be flattened 
	by just looking at the radial part of Higgs because there are NG bosons.
	In the gauged case, the longitudinal modes of the gauge bosons play 
	the same role as the NG bosons in the unitary gauge because of
	the NG boson equivalence theorem. 
	See Refs.~\cite{Burgess:2010zq,Hertzberg:2010dc,Burgess:2014lza,Ren:2014sya}.
}
In this frame, the light scalaron should stem from 
the property of this nontrivial kinetic term~\eqref{eq:nlsm-intro}.
Since physics is frame independent, it is desirable to understand the emergence of the scalaron in
a frame independent way.

Our main goal is thus to provide a frame independent understanding of 
Higgs inflation, the scalaron and its emergence.
To this end, we rewrite Higgs inflation as a nonlinear sigma model (NLSM).
A crucial point is that we include not only the Higgs field but also the conformal mode of the metric
in our definition of the NLSM. 
Here the conformal mode of the metric $\varphi$ is defined as
\begin{align}
	g_{\mu\nu} = e^{2\varphi} \tilde{g}_{\mu\nu},
\end{align}
with $\mathrm{det}[\tilde{g}_{\mu\nu}] = -1$ and $g_{\mu\nu}$ the spacetime metric.
The inclusion of the conformal mode is essential
since it provides us with a frame independent definition of the target space.
The large coupling $\xi$ controls the interaction between the conformal mode 
and the Higgs in the Jordan frame,
while it controls the interaction among the Higgs fields in the Einstein frame,
and both are equally captured by the geometry of our target space
which is invariant under the frame transformation.
Once written as the NLSM, one naturally expects a new scalar degree of freedom, \emph{$\sigma$-meson},
that linearizes the target space of Higgs inflation.
We see that this $\sigma$-meson is identified with the scalaron.
It UV-completes Higgs inflation to be a linear sigma model (LSM) with renormalizable interactions, 
consistent with the renormalizability of quadratic gravity~\cite{Weinberg:1974tw,Deser:1975nv,Stelle:1976gc,Barvinsky:2017zlx,Salvio:2018crh}.
Since our target space is frame independent,
this identification of the scalaron as the $\sigma$-meson is frame independent.

We then study quantum corrections of Higgs inflation in the large-$N$ limit. 
We show that a new scalar degree of freedom shows up in the spectrum 
which can be identified as the $\sigma$-meson and hence the scalaron,
naturally becoming light for $\xi \gg 1$.
We thus provide a frame independent understanding of the emergence of the scalaron
that was previously studied in a specific frame in Ref.~\cite{Ema:2019fdd}.
Formulated as a NLSM,
our large-$N$ analysis is clearly parallel to that of other models, such as
the O$(N)$ NLSM and the $\sigma$-meson~\cite{Bardeen:1976zh,Brezin:1976ap} 
(see Refs.~\cite{Dobado:1995qy,Dobado:1999xb} on the application to the Standard Model Higgs),
the CP$^{N-1}$ model that possesses the hidden local symmetry
and the $\rho$-meson~\cite{DAdda:1978vbw,DAdda:1978dle,Witten:1978bc,Arefeva:1980ms, Fujiwara:1984mp,Bando:1984ej,Bando:1987br},
and the Nambu--Jona-Lasinio and Gross--Neveu models 
and the scalar mesons~\cite{Nambu:1961tp,Nambu:1961fr,Gross:1974jv}.

Before moving to our main discussion, here we comment on other UV-completions of Higgs inflation
that are discussed in Refs.~\cite{Lerner:2010mq,Giudice:2010ka,Lee:2018esk,Barbon:2015fla}.
In particular, Ref.~\cite{Giudice:2010ka} emphasized
the importance of linearizing the Higgs kinetic term in the Einstein frame prior to us.
In this work, an additional scalar degree of freedom is added to UV-complete Higgs inflation
which is called a $\sigma$-field in analogy with the LSM.
There are two main differences between this work and ours.
First, the definition of the target space is different.
In Ref.~\cite{Giudice:2010ka}, the target space is defined solely by the kinetic terms of the scalar fields 
and hence it is frame dependent
as we will demonstrate below,
while our definition of the target space is frame independent.
Second, that target space is not completely flat even with the additional scalar
as the kinetic terms still contain (Planck-suppressed) higher dimensional operators.
On the other hand, the scalaron makes the target space
completely flat,
without any higher-dimensional operators in the scalar sector,
and can be identified as the $\sigma$-meson in a more strict sense.
Practically, however, the latter point may be less important
since our model also loses renormalizability at the Planck scale
at which the spin-2 graviton comes into play.

The organization of this paper is as follows.
In Sec.~\ref{sec:higgs_inf}, we rewrite Higgs inflation as a NLSM.
We include not only the Higgs but also the conformal mode of the metric in our definition of the target
space, and show that our definition is indeed frame independent.
In Sec.~\ref{sec:scalaron},
we show that the $\sigma$-meson of Higgs inflation is nothing but the scalaron.
In Sec.~\ref{sec:large-N}, we study quantum corrections of Higgs inflation in the large-$N$ limit,
and see that they give rise to the $\sigma$-meson, or the scalaron.
Finally, Sec.~\ref{sec:summary} is devoted to summary and discussion.

\section{Higgs inflation as NLSM}
\label{sec:higgs_inf}

In this section, we show that Higgs inflation can be interpreted as a NLSM.
Our primary goal of this section is to provide a frame independent definition of the target space.
For this purpose, we include not only the scalar fields but also the conformal mode of the metric
in the definition of the target space.

This section is composed of two parts.
In Sec.~\ref{subsec:target_space},
we show that Higgs inflation can be interpreted as a NLSM
with its target space given by
\begin{align}
	\frac{6 \xi + 1}{2} \phi_{i}^{2} + \left(h + \frac{\Phi}{2} \right)^{2} = \frac{\Phi^{2}}{4}
	\quad \text{in} \quad \left(\Phi, \phi_{i}, h \right) \in \mathbb{R}^{(1, N+1)},
	\label{eq:target}
\end{align}
where $\phi_{i}$ is a real scalar field (corresponding to each component of the Higgs doublet), 
$\Phi$ is the conformal mode of the metric (which we will define below), 
and $\xi$ is a nonminimal coupling.
The summation over the index $i$ ranging from $1$ to $N$ is implied, with $N = 4$ for Higgs inflation.\footnote{
	Note that the Higgs potential has a
	global symmetry under O$(4) \simeq$ SU$(2)_{\text{L}} \times \text{SU}(2)_{\text{R}}$,
	which leads to the custodial symmetry.
} In Sec.~\ref{subsec:frame_independence}, we show that our definition of the target space
is indeed frame independent.

We note here that we do not consider any quantum effects in this section.
Thus ``Higgs inflation" in this section always indicates the theory 
without any counter terms.
Quantum corrections of Higgs inflation are studied in the large-$N$ limit in Sec.~\ref{sec:large-N}.

\subsection{Target space of Higgs inflation}
\label{subsec:target_space}

We start from the action for Higgs inflation in the Jordan frame:
\begin{align}
	S = \int \dd^4 x \sqrt{-g_J}
	\left[\frac{M_P^2}{2}R_J\left(1 + \frac{\xi \phi_{Ji}^2}{M_P^2}\right)
	+ \frac{1}{2}g^{\mu\nu}_J \partial_\mu \phi_{Ji} \partial_\nu \phi_{Ji}
	- \frac{\lambda}{4}\left(\phi_{Ji}^2\right)^2
	\right],
	\label{eq:action_Jordan}
\end{align}
where $M_P$ is the reduced Planck mass, 
$g_{J\mu\nu}$ is the spacetime metric with $g_J$ its determinant, 
$R_J$ is the Ricci scalar,
$\xi$ is a nonminimal coupling,
$\lambda$ is the quartic coupling,
and $i = 1, ..., N$
with $N = 4$ for the Standard Model Higgs.
The subscript $J$ indicates that the quantities are defined in the Jordan frame.
In the following, we extract the conformal mode of the metric to define the target space of Higgs inflation 
frame independently.

Without loss of generality, the metric can be decomposed as
\begin{align}
	g_{J \mu\nu} = e^{2\varphi_J} \tilde{g}_{\mu\nu},
	\quad
	\mathrm{det}\left[ \tilde{g}_{\mu\nu} \right] = -1.
	\label{eq:metric_decom}
\end{align}
The scalar mode $\varphi_J$ contains the determinant part of the metric,
and we call it the conformal mode of the metric.
Note that the Weyl transformation
solely transforms the conformal mode $\varphi_J$, and not the other part $\tilde{g}_{\mu\nu}$,
and hence we do not put the index $J$ on $\tilde{g}_{\mu\nu}$.
We will come back to this point in Sec.~\ref{subsec:frame_independence}.
The Ricci scalar is then decomposed as
\begin{align}
	R_J = e^{-2\varphi_J} \tilde R + 6 e^{- 3 \varphi_J} \tilde \Box e^{\varphi_J},
	\label{eq:Ricci_decom}
\end{align}
where $\tilde R$ and $\tilde \Box$ are defined by $\tilde g_{\mu\nu}$.
We redefine the scalar fields as
\begin{align}
	\phi_{Ji} = e^{-\varphi_J} \phi_{i},
	\label{eq:redef}
\end{align}
and define
\begin{align}
	\Phi_J = \sqrt{6} M_P e^{\varphi_J},
	\label{eq:conf_tilde}
\end{align}
to which we also refer as the conformal mode.
As a result, we can rewrite the action as
\begin{align}
	S = \int \dd^4x &\left[
	\frac{\tilde R}{12} \left( \Phi_J^2 + 6 \xi \phi_i^2 \right)
	\right. \nonumber \\ &\left.
	-\frac{1}{2}\left(1-\left(6\xi + 1\right) \frac{\phi_i^2}{\Phi_J^2}\right) 
	\tilde{g}^{\mu\nu}\partial_\mu \Phi_J \partial_\nu \Phi_J
	+ \frac{1}{2}\tilde{g}^{\mu\nu}\partial_\mu \phi_i \partial_\nu \phi_i
	- \left(6\xi + 1\right) \frac{\phi_i}{\Phi_J} \tilde{g}^{\mu\nu} \partial_\mu \phi_i \partial_\nu \Phi_J
	- \frac{\lambda}{4}\left(\phi_i^2\right)^2
	\right].
	\label{eq:Higgs_NLSM_Jordan}
\end{align}
Thus, Higgs inflation is now written in the form of a NLSM composed of $\phi_i$ and $\Phi_J$.
It is, however, useful to move to a field basis 
in which the definition of the target space is more transparent.
For this purpose, we redefine the conformal mode as
\begin{align}
	\Phi_J = \frac{1}{2}\left[\sqrt{\Phi^2 - 2\left(6\xi + 1\right)\phi_i^2} + \Phi\right].
	\label{eq:Phi_redef}
\end{align}
The new field $\Phi$ satisfies
\begin{align}
	-\frac{6\xi + 1}{2}\frac{\phi_i^2}{\Phi_J} 
	= \frac{1}{2}\left[\sqrt{\Phi^2 - 2\left(6\xi + 1\right)\phi_i^2} - \Phi\right],
\end{align}
and hence the action~\eqref{eq:Higgs_NLSM_Jordan} is written in terms of $\Phi$ as 
\begin{align}
	S = \int \dd^4x \left[
	\frac{\tilde R}{12} \left( \Phi^2 - \phi_i^2 - h(\Phi, \phi)^2 \right)
	- \frac{1}{2}\tilde{g}^{\mu\nu}\partial_\mu \Phi \partial_\nu \Phi
	+ \frac{1}{2}\tilde{g}^{\mu\nu}\partial_\mu \phi_i \partial_\nu \phi_i
	+ \frac{1}{2}\tilde{g}^{\mu\nu}\partial_\mu h(\Phi, \phi) \partial_\nu h(\Phi, \phi)
	- \frac{\lambda}{4}\left(\phi_i^2\right)^2
	\right],
	\label{eq:Higgs_NLSM_cf}
\end{align}
where the scalar function $h(\Phi, \phi)$ is given by
\begin{align}
	h(\Phi, \phi) = \frac{1}{2}\left[ \sqrt{\Phi^2 - 2\left(6\xi + 1\right)\phi_i^2} - \Phi\right].
	\label{eq:h_target}
\end{align}
Here we again refer to this $\Phi$ as the conformal mode of the metric with a slight abuse of terminology.
Now it is clear that the target space of Higgs inflation is given by
\begin{align}
	\frac{6 \xi + 1}{2} \phi_{i}^{2} + \left(h + \frac{\Phi}{2} \right)^{2} = \frac{\Phi^{2}}{4}
	\quad \text{in} \quad \left(\Phi, \phi_{i}, h \right) \in \mathbb{R}^{(1, N+1)},
	\label{eq:target2}
\end{align}
which is an $N+1$-dimensional hypersurface in $\mathbb{R}^{(1,N+1)}$.
The curvature of the target space is controlled by the parameter~$6\xi + 1$.
In particular, if the scalar fields are conformally coupled to gravity, $\xi = -1/6$, the target space 
is flat and the action reduces to an LSM as expected.
An important feature of this target space is that
the kinetic term of $\Phi$ has the wrong sign,
and hence $\Phi$ is a ghost-like mode.
In fact, such a ghost exists even in pure Einstein gravity 
(see, \emph{e.g.}, Ref.~\cite{Alvarez-Gaume:2015rwa}),
which resembles the time-like component of the U(1) gauge field in the Lorenz gauge.
Although ghost-like, it is harmless 
thanks to a residual gauge symmetry. See App.~\ref{sec:gauge} for more details on this point.

In Sec.~\ref{subsec:frame_independence},
we show that our definition of the target space is indeed frame independent
thanks to the inclusion of the conformal mode of the metric.

\subsection{Frame independence of target space}
\label{subsec:frame_independence}
In this subsection, we show that our definition of the target space is frame independent.
Before going to our main discussion, however, let us first emphasize that a naive definition of the target space
solely by the kinetic terms of the scalar fields is frame dependent.
For instance, the action for Higgs inflation in the Jordan frame is given by Eq.~\eqref{eq:action_Jordan},
and hence the kinetic terms of the scalar fields are completely flat in this frame.
Once we move to, \textit{e.g.}, the Einstein frame by
\begin{align}
	g_{J\mu\nu} = \Omega_E^{-2} g_{E\mu\nu},
	\quad
	\Omega_E^2 = 1 + \frac{\xi \phi_{Ji}^2}{M_P^2},
\end{align}
the action is given by
\begin{align}
	S = \int \dd^4 x \sqrt{-g_E}
	\left\{
	\frac{M_P^2}{2} R_E 
	+ \frac{1}{2\Omega_E^4}
	\left[ \left(1+ \frac{\xi \phi_{Jk}^2}{M_P^2}\right)\delta_{ij} + \frac{6\xi^2 \phi_{Ji} \phi_{Jj}}{M_P^2}\right]
	g^{\mu\nu}_E\partial_\mu \phi_{Ji} \partial_\nu \phi_{Jj}
	- \frac{\lambda \left( \phi_{Ji}^2 \right)^2}{4\Omega_E^4} 
	\right\}.
	\label{eq:Higgs_Einstein_frame_indep}
\end{align}
The kinetic terms of the scalar fields are now more involved, and one cannot canonically normalize 
all the scalar fields at the same time (unless there is only one real scalar field, or $N = 1$).
As a result, the kinetic terms of the scalar fields are curved in this frame,
and hence the definition of the target space based solely on the kinetic terms of the scalar fields
is frame dependent.
Since physics such as the unitarity violation scale is frame independent,
it is desirable to define the target space in a frame independent way.

Now we show that our definition of the target space is frame independent.
The frame transformation, or the Weyl transformation, from a frame $A$ to a frame $B$
is given by
\begin{align}
	g_{A\mu\nu} = \Omega^{-2}g_{B\mu\nu},
\end{align}
with some function $\Omega$.
With the metric decomposition~\eqref{eq:metric_decom}, 
it can be written as a field redefinition of the conformal mode,
\begin{align}
	\Phi_A^2 = \Omega^{-2} \Phi_B^2,
\end{align}
where
\begin{align}
	g_{\bullet \mu\nu} = e^{2\varphi_\bullet} \tilde g_{\mu\nu}, \quad
	\Phi_\bullet \equiv \sqrt{6} M_P e^{\varphi_\bullet},
\end{align}
with $\bullet = A, B$.
Thus, the frame transformation is a particular form of a coordinate transformation of 
our target space since we include the conformal mode as a coordinate.
The frame independence of our target space immediately
follows since the target space is in general invariant under a coordinate transformation
(see, \textit{e.g.}, Ref.~\cite{Ketov:2000dy}).
It means that geometrical quantities such as the curvature of the target space are not affected
by the frame transformation.
One can also see that the curvature of the target space in our definition 
is directly translated to the cut-off scale of Higgs inflation by, \textit{e.g.}, computing scattering amplitudes,
implying that our definition of the target space is of physical importance.

Although the above argument already proves the frame independence of our target space,
it may be instructive to see what is going on in more detail with an example.
For this reason, 
we consider the frame transformation between the Jordan and Einstein frames
in the following.
We explicitly write down a field redefinition among the conformal mode and the scalar fields that 
corresponds to a frame transformation in this case.

The actions for Higgs inflation in the Jordan and Einstein frames are respectively given by
Eqs.~\eqref{eq:action_Jordan} and~\eqref{eq:Higgs_Einstein_frame_indep}.
By extracting the conformal modes,
the action in the Jordan frame is given by Eq.~\eqref{eq:Higgs_NLSM_Jordan},
while that in the Einstein frame is given by
\begin{align}
	S = \int \dd^4 x &\left\{
	\frac{\tilde{R}}{12}\Phi_E^2
	- \frac{1}{2}\tilde{g}^{\mu\nu}\partial_\mu \Phi_E \partial_\nu \Phi_E
	\right. \nonumber \\ &\left.
	+ \frac{\Phi_E^2}{12M_P^2 \Omega_E^4}
	\left[\left(1+\frac{\xi \phi_{Jk}^2}{M_P^2}\right)\delta_{ij}
	+ \frac{6\xi^2 \phi_{Ji} \phi_{Jj}}{M_P^2} \right]
	\tilde{g}^{\mu\nu}\partial_\mu \phi_{Ji} \partial_\nu \phi_{Jj}
	- \frac{\lambda \Phi_E^4}{144M_P^4 \Omega_E^4}\left(\phi_{Ji}^2\right)^2
	\right\},
	\label{eq:Higgs_NLSM_Einstein}
\end{align}
where
\begin{align}
	g_{E\mu\nu} = e^{2\varphi_E}\tilde{g}_{\mu\nu},
	\quad
	\Phi_E = \sqrt{6}M_Pe^{\varphi_E}.
\end{align}
We can move back and forth between these two actions by redefining
the conformal mode and the scalar fields as
\begin{align}
	\Phi_E^2 = \Phi_J^2 + 6\xi \phi_i^2,
	\quad
	\phi_{Ji} = e^{-\varphi_E}\Omega_E \phi_i.
	\label{eq:JtoE}
\end{align}
Indeed, this redefinition implies that
\begin{align}
	\frac{\Phi_E^2}{12M_P^2\Omega_E^2}\tilde{g}^{\mu\nu}\partial_\mu \phi_{Ji} \partial_\nu \phi_{Ji}
	&= \frac{1}{2}\tilde{g}^{\mu\nu}\partial_\mu \phi_i \partial_\nu \phi_i
	+ \frac{\phi_i^2}{2\Phi_J^2}\tilde{g}^{\mu\nu}\partial_\mu \Phi_J \partial_\nu \Phi_J
	- \frac{\phi_i}{\Phi_J}\tilde{g}^{\mu\nu}\partial_\mu \phi_i \partial_\nu \Phi_J, \\
	-\frac{1}{2}\tilde{g}^{\mu\nu}\partial_\mu \Phi_E \partial_\nu \Phi_E
	+ \frac{\Phi_E^2}{8M_P^4 \Omega_E^4}\tilde{g}^{\mu\nu} \partial_\mu \phi_{Ji}^2 \partial_\nu \phi_{Jj}^2
	&= 
	-\frac{1}{2}\left(1 - \frac{6\xi \phi_i^2}{\Phi_J^2}\right)\tilde{g}^{\mu\nu} \partial_\mu \Phi_J \partial_\nu \Phi_J
	- \frac{6\xi \phi_i}{\Phi_J}\tilde{g}^{\mu\nu} \partial_\mu \phi_i \partial_\nu \Phi_J,
\end{align}
up to total derivative,
and one can recover Eq.~\eqref{eq:Higgs_NLSM_Jordan}
from Eq.~\eqref{eq:Higgs_NLSM_Einstein} by inserting these expressions.
This confirms that the frame transformation between the Jordan and the Einstein frames 
corresponds to the redefinition of the conformal mode given in Eq.~\eqref{eq:JtoE}.

It is also instructive to see to which frame the conformal mode $\Phi$ in Eq.~\eqref{eq:Higgs_NLSM_cf}
corresponds.
Let us define a metric by
\begin{align}
	\sqrt{6} M_P e^{\varphi_C} \equiv \Phi, \quad
	g_{C\mu\nu} \equiv e^{\varphi_C} \tilde g_{\mu\nu}.
\end{align}
By redefining the fields as
\begin{align}
	\phi_i = e^{\varphi_C} \phi_{Ci},
\end{align}
we easily obtain the following action
\begin{align}
	S = \int \dd^4 x \sqrt{-g_C}
	&\left[
	\frac{R_C}{12}\left(6M_P^2- \phi_{Ci}^2 - h_C^2\right)
	+ \frac{1}{2}g_C^{\mu\nu} \partial_\mu \phi_{Ci} \partial_\nu \phi_{Ci}
	+ \frac{1}{2}g_C^{\mu\nu} \partial_\mu h_C \partial_\nu h_C
	- \frac{\lambda}{4}\left(\phi_{Ci}^2 \right)^2
	\right],
	\label{eq:action_cf}
\end{align}
where the scalar function $h_C = h_C(\phi_C)$ is given by
\begin{align}
	h_C = \frac{1}{2}\left[
	\sqrt{6M_P^2-2\left(6\xi+1\right)\phi_{Ci}^2} - \sqrt{6}M_P
	\right].
\end{align}
This expression describes Higgs inflation in the conformal frame.
Of course Eq.~\eqref{eq:action_cf} can be derived 
directly from Eq.~\eqref{eq:action_Jordan} by the Weyl transformation.
This confirms that the frame transformation 
from the Jordan frame to the conformal frame
corresponds to the field redefinition of the conformal mode~\eqref{eq:Phi_redef}.

\section{Scalaron as $\sigma$-meson}
\label{sec:scalaron}

In Sec.~\ref{sec:higgs_inf}, we have shown that Higgs inflation
can be regarded as 
a NLSM on an $N+1$-dimensional hypersurface spanned by the Higgs $\phi_{i}$ 
and the conformal mode of the metric $\Phi$ in $\mathbb{R}^{1,N+1}$.
This structure can be seen easily in a particular basis as shown in Eq.~\eqref{eq:Higgs_NLSM_cf}:
\begin{align}
	S = \int \dd^4x \left[
	\frac{\tilde R}{12} \left( \Phi^2 - \phi_i^2 - h^2 \right)
	- \frac{1}{2}\tilde{g}^{\mu\nu}\partial_\mu \Phi \partial_\nu \Phi
	+ \frac{1}{2}\tilde{g}^{\mu\nu}\partial_\mu \phi_i \partial_\nu \phi_i
	+ \frac{1}{2}\tilde{g}^{\mu\nu}\partial_\mu h \partial_\nu h
	- \frac{\lambda}{4}\left(\phi_i^2\right)^2
	\right],
	\label{eq:Higgs_NLSM_cf_2}
\end{align}
where
\begin{align}
	h = \frac{1}{2}\left[ \sqrt{\Phi^2 - 2\left(6\xi + 1\right)\phi_i^2} - \Phi\right].
\end{align}
Because of its simple form,
one can naturally linearize and hence UV-complete this NLSM 
by promoting $h$ to a fundamental field as
\begin{align}
	S = \int \dd^{4}x\,
	&\left\{
	\frac{\tilde R}{12} \left( \Phi^2 - \phi_i^2 - \sigma^2 \right)
	- \frac{1}{2}\tilde{g}^{\mu\nu}\partial_\mu \Phi \partial_\nu \Phi
	\right. \nonumber \\ &\left.
	+ \frac{1}{2}\tilde{g}^{\mu\nu}\partial_\mu \phi_i \partial_\nu \phi_i
	+ \frac{1}{2}\tilde{g}^{\mu\nu}\partial_\mu \sigma \partial_\nu \sigma
	- \frac{\lambda}{4}\left(\phi_i^2\right)^2
	- \frac{1}{144 \alpha} \left[
		\frac{\Phi^{2}}{4} - \left(\sigma + \frac{\Phi}{2} \right)^{2} - \frac{6 \xi + 1}{2} \phi_{i}^{2}
		\right]^{2}
	\right\}.
	\label{eq:Higgs_LSM_cf}
\end{align}
One can see that it goes back to the original NLSM~\eqref{eq:Higgs_NLSM_cf_2}
in the limit $\alpha \to 0$.
We denote the additional field by $\sigma$ since 
it completely linearizes the target space and hence corresponds to the $\sigma$-meson
in the language of the NLSM.
We emphasize that the notions of the flatness of the target space and hence the $\sigma$-meson
are frame independent since our definition of the target space is frame independent.

The primary goal of this section is to show that this $\sigma$-meson is nothing but the scalaron
that arises due to the $R^2$~term in the Jordan frame.
We also comment on the unitarity and renormalizability of the resultant LSM~\eqref{eq:Higgs_LSM_cf}.

\subsection{Scalaron as $\sigma$-meson}
Since the scalaron is understood in the literature as the degree of freedom that originates from 
the $R^2$~term in the Jordan frame,
we start from the following action,
\begin{align}
	S = \int \dd^4 x \sqrt{-g_J}
	\left[\frac{M_P^2}{2}R_J\left(1 + \frac{\xi \phi_{Ji}^2}{M_P^2}\right)
	+ \alpha R_J^2
	+ \frac{1}{2}g^{\mu\nu}_J \partial_\mu \phi_{Ji} \partial_\nu \phi_{Ji}
	- \frac{\lambda}{4}\left(\phi_{Ji}^2\right)^2
	\right],
\end{align}
and show that it coincides with the LSM~\eqref{eq:Higgs_LSM_cf} by appropriate field redefinitions.
It shows that the scalaron can be identified with the $\sigma$-meson
that linearizes Higgs inflation.

As before, we extract the conformal mode of the metric as
\begin{align}
	g_{J \mu\nu} = e^{2\varphi_J} \tilde{g}_{\mu\nu},
	\quad
	\mathrm{det}\left[ \tilde{g}_{\mu\nu} \right] = -1,
\end{align}
and redefine the fields as
\begin{align}
	\phi_{Ji} = e^{-\varphi_J}\phi_i,
	\quad
	\Phi_J = \sqrt{6}M_P e^{\varphi_J}.
\end{align}
The action is then given by
\begin{align}
	S = \int \dd^4 x 
	&\left[
	\frac{\tilde{R}}{12}\left(\Phi_J^2 + 6\xi \phi_i^2\right)
	- \frac{1}{2}\tilde{g}^{\mu\nu}\partial_\mu \Phi_J \partial_\nu \Phi_J
	\right. \nonumber \\ &\left.
	+ \frac{1}{2}\tilde{g}^{\mu\nu}\partial_\mu \phi_i \partial_\nu \phi_i
	+ \left(\frac{6\xi+1}{2}\phi_i^2 + 12\alpha \tilde{R}\right)\frac{\tilde{\Box} \Phi_J}{\Phi_J}
	+ \alpha \tilde{R}^2 + 36\alpha \left(\frac{\tilde{\Box}\Phi_J}{\Phi_J}\right)^2
	- \frac{\lambda}{4}\left(\phi_i^2\right)^2
	\right].
	\label{eq:Higgs_LSM_Jordan}
\end{align}
Since it contains the higher derivative term, $(\tilde{\Box} \Phi_J/\Phi_J)^2$,
it contains an additional degree of freedom, corresponding to the scalaron.
We extract it by adding an auxiliary field $\sigma_J$ as
\begin{align}
	S = \int \dd^4 x 
	&\left\{
	\frac{\tilde{R}}{12}\left(\Phi_J^2 + 6\xi \phi_i^2\right)
	- \frac{1}{2}\tilde{g}^{\mu\nu}\partial_\mu \Phi_J \partial_\nu \Phi_J
	+ \frac{1}{2}\tilde{g}^{\mu\nu}\partial_\mu \phi_i \partial_\nu \phi_i
	\right. \nonumber \\ &\left.
	+ \left(\frac{6\xi+1}{2}\phi_i^2 + 12\alpha \tilde{R}\right)\frac{\tilde{\Box} \Phi_J}{\Phi_J}
	+ \alpha \tilde{R}^2 + 36\alpha 
	\left[\left(\frac{\tilde{\Box}\Phi_J}{\Phi_J}\right)^2
	- \left(\frac{\tilde{\Box}\Phi_J}{\Phi_J} + \frac{\Phi_J \sigma_J}{72\alpha}\right)^2\right]
	- \frac{\lambda}{4}\left(\phi_i^2\right)^2
	\right\}.
\end{align}
It is obvious that it reduces to the original action by integrating out $\sigma_J$.
By further defining the fields as
\begin{align}
	\sigma_J = \sigma + \frac{6\xi+1}{2}\frac{\phi_i^2}{\Phi_J} + 12\alpha \frac{\tilde{R}}{\Phi_J},
	\quad
	\Phi_J = \Phi + \sigma,
\end{align}
we arrive at our final result
\begin{align}
	S = \int \dd^{4}x\,
	&\left\{
	\frac{\tilde R}{12} \left( \Phi^2 - \phi_i^2 - \sigma^2 \right)
	- \frac{1}{2}\tilde{g}^{\mu\nu}\partial_\mu \Phi \partial_\nu \Phi
	\right. \nonumber \\ &\left.
	+ \frac{1}{2}\tilde{g}^{\mu\nu}\partial_\mu \phi_i \partial_\nu \phi_i
	+ \frac{1}{2}\tilde{g}^{\mu\nu}\partial_\mu \sigma \partial_\nu \sigma
	- \frac{\lambda}{4}\left(\phi_i^2\right)^2
	- \frac{1}{144 \alpha} \left[
		\frac{\Phi^{2}}{4} - \left(\sigma + \frac{\Phi}{2} \right)^{2} - \frac{6 \xi + 1}{2} \phi_{i}^{2}
		\right]^{2}
	\right\},
\end{align}
which coincides with Eq.~\eqref{eq:Higgs_LSM_cf}.
Thus, we have shown that the scalaron is nothing but the $\sigma$-meson
that linearizes the target space of Higgs inflation.

It may be instructive to rewrite the action in the form before extracting the conformal mode.
Let us define a metric by
\begin{align}
	\sqrt{6} M_P e^{\varphi_C} \equiv \Phi, \quad
	g_{C\mu\nu} \equiv e^{\varphi_C} \tilde g_{\mu\nu},
\end{align}
and redefine the fields as
\begin{align}
	\phi_i = e^{\varphi_C} \phi_{Ci},
	\quad
	\sigma = e^{\varphi_C} \sigma_C.
\end{align}
We then obtain
\begin{align}
	S = \int \dd^{4}x\sqrt{-g_C}\,
	&\left\{
	\frac{R}{12} \left(6M_P^2 - \phi_i^2 - \sigma^2 \right)
	+ \frac{1}{2}{g}^{\mu\nu}\partial_\mu \phi_{Ci} \partial_\nu \phi_{Ci}
	\right. \nonumber \\ &\left.
	+ \frac{1}{2}{g}^{\mu\nu}\partial_\mu \sigma_C \partial_\nu \sigma_C
	- \frac{\lambda}{4}\left(\phi_{Ci}^2\right)^2
	- \frac{1}{144 \alpha} \left[
		\frac{3M_P^2}{2} - \left(\sigma_C + \frac{\sqrt{6}M_P}{2} \right)^{2} - \frac{6 \xi + 1}{2} \phi_{Ci}^{2}
		\right]^{2}
	\right\}.
\end{align}
It describes the Higgs-scalaron system in the conformal frame.
Thus the flatness of our target space corresponds to the flatness of the kinetic terms
of the scalar fields in the conformal frame.

Here is one remark. Additional degrees of freedom that arise 
due to higher derivative terms are often ghost-like, known as Ostrogradsky ghosts
(see, \emph{e.g.}, Ref.~\cite{Woodard:2006nt} and references therein).
In our case, however, $\sigma$ 
has a kinetic term with the correct sign, and hence is healthy.
It is because $\Phi$ has a kinetic term 
with the wrong sign and is ghost-like. 
Thus, we may phrase this phenomenon as ``minus times minus gives plus,"  or ``the ghost of a ghost is healthy." 

\subsection{Unitarity and renormalizability}
\label{sec:unit-ren}

We have seen that, if we regard Higgs inflation 
as an NLSM~\eqref{eq:Higgs_NLSM_cf},
 the scalaron is understood as the $\sigma$-meson which UV-completes it to an LSM~\eqref{eq:Higgs_LSM_cf}.
A remarkable feature of the LSM~\eqref{eq:Higgs_LSM_cf} is that it has a completely flat target space 
and its scalar potential involves only terms that are quartic in the fields.
It indicates that 
the Higgs-scalaron system can be unitary and renormalizable up to a very high energy scale 
as far as it does not hit a Landau pole.
Indeed, an explicit computation shows that the LSM~\eqref{eq:Higgs_LSM_cf}
with the inclusion of the Higgs mass term and the cosmological constant is renormalizable
even above the Planck scale as far as the scalar sector is concerned
(by taking $\tilde{g}_{\mu\nu} = \eta_{\mu\nu}$).
It is consistent with the analysis based on the scattering amplitude in Ref.~\cite{Ema:2019fdd}.
In reality, of course, the renormalizability is lost by the presence of the spin-$2$ graviton.
Still, the field basis given in Eq.~\eqref{eq:Higgs_LSM_cf} is useful for computing important quantities such as the quantum corrections and the RG running of the potential 
up to the energy scale where
the spin-2 graviton comes into play (which corresponds to the Planck scale in the Einstein frame).
See Ref.~\cite{Ema:2020evi} for more details on this point.

We note that these properties correspond to the renormalizability of quadratic gravity~\cite{Weinberg:1974tw,Deser:1975nv,Stelle:1976gc,Barvinsky:2017zlx,Salvio:2018crh}.
As far as the scalar sector is concerned, the Higgs-scalaron system is equivalent to quadratic gravity with scalar fields nonminimally coupled to gravity, since the other operator in quadratic gravity, 
$R_{\mu\nu}R^{\mu\nu} - R^2/3$, only affects the tensor sector, 
leading to the infamous spin-$2$ ghost. 
Hence, the unitarity and renormalizability of quadratic gravity up to the Planck scale can also be understood 
as a property of Eq.~\eqref{eq:Higgs_LSM_cf}.
Although other field bases such as Eq.~\eqref{eq:Higgs_LSM_Jordan} are equivalent to Eq.~\eqref{eq:Higgs_LSM_cf}, properties such as the unitarity scale and renormalizabitily are 
more difficult to see in these other bases.
The power of Eq.~\eqref{eq:Higgs_LSM_cf} comes from its appropriate field basis which makes the flatness of the target space manifest.

\section{Large-$N$ analysis of Higgs inflation}
\label{sec:large-N}

In this section, we study quantum corrections to Higgs inflation in the large-$N$ limit.
Here $N$ is the number of the real scalar fields, and the SM Higgs corresponds to $N = 4$.
In this section, we focus on the conformal mode of the metric and drop the spin-2 sector of the metric
by assuming $\xi \gg 1$.
In Sec.~\ref{subsec:mode_decomposition}, we explain why $\xi \gg 1$ allows us to ignore the spin-2 sector.
Then in Sec.~\ref{subsec:higgs_inf}, we study quantum corrections to Higgs inflation in the large-$N$ limit.
There we see that an additional degree of freedom emerges
that linearizes the target space completely and hence is identified as the scalaron.
We end this section with some remarks on the large-$N$ analysis in Sec.~\ref{subsec:remarks_largeN}.

Let us again emphasize that the emergence of 
a new degree of freedom from quantum corrections
is not unique to our NLSM. 
See the introduction for concrete examples of other models in which this happens.
It is part of the virtue of mapping Higgs inflation to the NLSM that we can see the similarity between the analysis in this paper (and Refs.~\cite{Aydemir:2012nz,Calmet:2013hia,Ema:2019fdd}) and the literature referred to 
in the introduction.

\subsection{Mode decomposition}
\label{subsec:mode_decomposition}

In this section, we assume that $\xi \gg 1$ and hence ignore the spin-2 sector of gravity.
In order to see why the limit $\xi \gg 1$ allows us to ignore the spin-2 sector,
we study the interaction between the metric and the matter fields in this subsection.

In order to study the interaction between $\tilde{g}_{\mu\nu}$ and the matter fields,
we may expand $\tilde{g}_{\mu\nu}$ around a flat spacetime metric as
\begin{align}
	\tilde{g}_{\mu\nu} = \eta_{\mu\nu} + h_{\mu\nu},
\end{align}
where $\eta_{\mu\nu} = \mathrm{diag}(1,-1,-1,-1)$ is the flat spacetime metric.
A small perturbation $h_{\mu\nu}$ can be further decomposed as
\begin{align}
	h_{\mu\nu} = h_{\mu\nu}^{\perp} + \partial_{\mu} h_{\nu}^{\perp} + \partial_{\nu} h_{\mu}^{\perp}
	+ \left(\partial_{\mu}\partial_{\nu} - \frac{1}{4}\eta_{\mu\nu}\Box\right)\psi.
	\label{eq:decomp_gravity}
\end{align}
The modes $h^{\perp}_{\mu\nu}$ and $h^{\perp}_{\mu}$ satisfy
\begin{align}
	{h^{\perp \mu}}_{\mu} = \partial^{\mu}h_{\mu\nu}^{\perp} = 0,
	\quad
	\partial^{\mu}h^{\perp}_{\mu} = 0,
\end{align}
where the contractions are taken by $\eta_{\mu\nu}$.
Note that $h_{\mu\nu}$ is traceless, $\eta^{\mu\nu} h_{\mu\nu} = 0$, since the determinant of $\tilde{g}_{\mu\nu}$ is unity.
Thus, before imposing any gauge fixing conditions, the metric contains 
one tensor mode $h^{\perp}_{\mu\nu}$ (five components), one vector mode $h^{\perp}_{\mu}$ (three components), 
and two scalar modes $\psi$ and $\varphi$,\footnote{
Here the words "scalar/vector/tensor" are defined under the Lorentz transformation 
as in the standard quantum field theory language.
They should not be confused with the scalar/vector/tensor decomposition in the context of the cosmological perturbation,
since the latter is defined only under the spatial rotation, not under the full Lorentz transformation.
}
and has in total ten components.
We can eliminate some of these components by a general coordinate transformation.
We may take the gauge fixing condition (at first order in perturbations) as
\begin{align}
	\partial^\mu h_{\mu\nu} = 0,
\end{align}
which kills $h^{\perp}_{\mu}$ and $\psi$.
Actually such a gauge fixing condition leaves a residual gauge symmetry,
which makes $\varphi$ and three out of five components in $h^{\perp}_{\mu\nu}$ unphysical,
resulting in two physical degrees of freedom (corresponding to two polarizations of the tensor mode).
Nevertheless, the conformal mode is crucial for our discussion since, 
although not dynamical, it still contributes to the scattering amplitude 
and hence the unitarity structure of the theory.\footnote{
	It is the same as a scattering of electrons; 
	the coulomb potential is not dynamical,
	yet contributes to the scattering.
}
See App.~\ref{sec:gauge} for more details on the residual gauge symmetry.

We now consider the coupling between the remaining modes $h^{\perp}_{\mu\nu}$ and $\varphi$, and the scalar fields.
The action of Higgs inflation in the Jordan frame is given by
\begin{align}
	S = \int \dd^4 x \sqrt{-g}
	\left[\frac{M_P^2}{2}R\left(1 + \frac{\xi \phi_i^2}{M_P^2}\right)
	+ \frac{1}{2}g^{\mu\nu} \partial_\mu \phi_i \partial_\nu \phi_i
	- \frac{\lambda}{4}\left(\phi_i^2\right)^2
	\right].
\end{align}
The stress energy tensor in flat spacetime is constructed from this action as
\begin{align}
	T_{\mu\nu}
	&= \left.\frac{2}{\sqrt{-g}}\frac{\delta S_\mathrm{matter}}{\delta g_{\mu\nu}}\right\rvert_{g_{\mu\nu} = \eta_{\mu\nu}} \nonumber \\
	&= \partial_\mu \phi_i \partial_\nu \phi_i 
	- \eta_{\mu\nu}\left(\frac{1}{2}\eta^{\alpha\beta}\partial_\alpha \phi_i \partial_\beta \phi_i - \frac{\lambda}{4}\phi_i^4\right)
	+ \xi \left(\partial_\mu \partial_\nu - \eta_{\mu\nu}\Box\right)\phi_i^2.
	\label{eq:em_tensor}
\end{align}
Since the metric couples to the stress energy tensor,
it follows that the interaction between $h^{\perp}_{\mu\nu}$ and the Higgs 
is independent of $\xi$: 
\begin{align}
	h^{\perp}_{\mu\nu} T^{\mu\nu}
	= h^{\perp}_{\mu\nu} \partial^\mu \phi_i \partial^\nu \phi_i,
\end{align}
up to total derivatives, where we have used the transverse-tracelessness of $h^{\perp}_{\mu\nu}$.
On the other hand, the interaction between $\varphi$ and the Higgs depends on $\xi$, 
as one can see, \textit{e.g.}, by taking the trace of the above stress energy tensor.
It means that the coupling to $h^{\perp}_{\mu\nu}$ is suppressed by $M_P$
whereas that to $\varphi$ is suppressed only by $M_P/\xi$
(after canonically normalizing the modes).
This is the reason why the $R^2$ operator, originating from the coupling to $\varphi$, appears at $M_P/\xi$ while $R_{\mu\nu} R^{\mu\nu} - R^2 / 3$, originating from the coupling to $h^\perp_{\mu\nu}$ appears at $M_P$.
Thus we focus on the conformal mode of the metric $\varphi$ with the assumption $\xi \gg 1$ in this section.
Here we have discussed the interaction in the Jordan frame,
but the fact that the interaction between the spin-2 sector and the matter fields is suppressed by $M_P$
is of course independent of the frame choice.

\subsection{Emergence of $\sigma$-meson as scalaron}
\label{subsec:higgs_inf}
As we have discussed in Sec.~\ref{subsec:mode_decomposition},
we drop the spin-2 sector in this subsection.
This is valid as long as $\xi \gg 1$ and the energy scale of our interest is below the Planck scale.
By taking
\begin{align}
	\tilde{g}_{\mu\nu} = \eta_{\mu\nu},
\end{align}
in the action~\eqref{eq:Higgs_NLSM_Jordan},
we thus obtain 
\begin{align}
	S  = \int \dd^4x \left[
	- \frac{1}{2}\left(\partial \Phi_J \right)^2
	+ \frac{1}{2}\left(\partial \phi_i \right)^2 
	+ \frac{6\xi+1}{2}\left(\frac{\Box \Phi_J}{\Phi_J}\right) \phi_i^{2} 
	- \frac{\lambda}{4} \left( \phi_{i}^{2} \right)^{2}
	\right].
	\label{eq:nlsm-cl-higgs}
\end{align}
The contraction of the Lorentz indices is always taken by $\eta_{\mu\nu}$ in this subsection.
We study quantum effects of this model in the large-$N$ limit.
These quantum effects induce divergences 
that have to be renormalized by counter terms.
Our primary goal of this subsection is to study what sort of divergences appear and 
what sort of counter terms are required to renormalize them in Higgs inflation 
at the leading order in the large-$N$ limit.
Here we keep the Higgs four-point interaction to clarify its effect in the large-$N$ analysis.

Let us first focus on divergences involving the Higgs four-point interaction.
Adopting dimensional regularization, we have two divergent diagrams in the large-$N$ limit
\begin{align}
	\begin{tikzpicture}[baseline=(c)]
	\begin{feynman}[inline = (base.c), horizontal=d to e]
		\vertex (c);
		\vertex [above = 1.2em of c] (i1);
		\vertex [below = 1.2em of c] (i2);
		\vertex [right = of c] (d);
		\vertex [right = of d] (e);
		\vertex [right = of e] (f);
		\vertex [above = 1.2em of f] (f2);
		\vertex [below = 1.2em of f] (f3);
		\diagram*{
		(i1) -- [scalar] (d),
		(i2) -- [scalar] (d),
		(e) -- [scalar, half left] (d) -- [scalar, half left] (e),
		(e) -- [scalar] (f2),
		(e) -- [scalar] (f3)
		};
	\end{feynman}
	\end{tikzpicture}
	\qquad
	\begin{tikzpicture}[baseline=(c)]
	\begin{feynman}[inline = (base.c), horizontal=d to e]
		\vertex (c);
		\vertex [right = of c] (d);
		\vertex [right = of d] (e);
		\vertex [right = of e] (f);
		\vertex [above = 1.2em of f] (f2);
		\vertex [below = 1.2em of f] (f3);
		\diagram*{
		(c) -- (d),
		(e) -- [scalar, half left] (d) -- [scalar, half left] (e),
		(e) -- [scalar] (f2),
		(e) -- [scalar] (f3)
		};
	\end{feynman}
	\end{tikzpicture}
\end{align}
where the solid line indicates the operator $\Box \Phi_J/ \Phi_J$ and the dotted line the scalar fields $\phi_i$.
The first diagram is renormalized by the Higgs four-point coupling, and the second one is renormalized by the nonminimal coupling $6\xi + 1$.
Hence, to cure the divergences involving the Higgs four-point interaction, we do not need to introduce any additional operators, since both of them are already present in Eq.~\eqref{eq:nlsm-cl-higgs}.
It is straightforward to check that they correctly reproduce the running of the Higgs four-point coupling and the nonminimal coupling.

On the other hand, a new operator is required in order to renormalize the two-point function of the operator $\Box \Phi_J/\Phi_J$, which is diagrammatically given by
\begin{align}
	\begin{tikzpicture}[baseline=(c)]
	\begin{feynman}[inline]
		\vertex (c);
		\vertex [right = of c] (d);
		\vertex [right = of d] (e);
		\vertex [right = of e] (f);
		\diagram*{
		(c) -- (d),
		(e) -- [scalar, half left] (d) -- [scalar, half left] (e),
		(e) -- (f)
		};
	\end{feynman}
	\end{tikzpicture}
\end{align}
and whose corresponding counter term is 
\begin{align}
	\mathcal{L}_\mathrm{c.t.} = 36\alpha \left(\frac{\Box \Phi_J}{\Phi_J}\right)^2.
	\label{eq:NLSM_higgs_ct}
\end{align}
Note that the divergences at the higher loop level, which are diagrammatically given by 
\begin{align}
	\begin{tikzpicture}[baseline=(c)]
	\begin{feynman}[inline = (base.c)]
		\vertex (c);
		\vertex [right = of c] (d);
		\vertex [right = of d] (e);
		\vertex [right = of e] (f);
		\vertex [right = of f] (g);
		\vertex [right = of g] (h);
		\diagram*{
		(c) -- (d),
		(e) -- [scalar, half left] (d) -- [scalar, half left] (e),
		(e) -- (f),
		(g) -- [scalar, half left] (f) -- [scalar, half left] (g),
		(g) -- (h)
		};
	\end{feynman}
	\end{tikzpicture}
	+
	\begin{tikzpicture}[baseline=(c)]
	\begin{feynman}[inline]
		\vertex (c);
		\vertex [right = of c] (d);
		\vertex [right = of d] (e);
		\vertex [right = of e] (f);
		\vertex [right = of f] (g);
		\vertex [right = of g] (h);
		\vertex [right = of h] (i);
		\vertex [right = of i] (j);
		\diagram*{
		(c) -- (d),
		(e) -- [scalar, half left] (d) -- [scalar, half left] (e),
		(e) -- (f),
		(g) -- [scalar, half left] (f) -- [scalar, half left] (g),
		(g) -- (h),
		(i) -- [scalar, half left] (h) -- [scalar, half left] (i),
		(i) -- (j)
		};
	\end{feynman}
	\end{tikzpicture}
	+
	\cdots
\end{align}
are renormalized by the same term~\eqref{eq:NLSM_higgs_ct},
and hence no other terms are required at the leading order in the large-$N$ limit.
We obtain the RG running of $\alpha$
\begin{align}
	\frac{\dd\alpha}{\dd\ln \mu} = - \frac{N}{1152\pi^2}\left(6\xi+1\right)^2,
	\label{eq:alphaRG_higgs}
\end{align}
which coincides with the running of the $R^{2}$ term Eq.~\eqref{eq:jordan-running}.
The value of $\alpha$ at a specific energy scale depends on the boundary condition which is a parameter choice of the theory.
Including quantum corrections at the leading order in the large-$N$ limit, the classical action \eqref{eq:nlsm-cl-higgs} is now modified to
\begin{align}
	S  = \int \dd^4x \left[
	- \frac{1}{2}\left(\partial \Phi_J\right)^2
	+ \frac{1}{2}\left(\partial \phi_i \right)^2 
	+ \frac{6\xi + 1}{2}\left(\frac{\Box \Phi_J}{\Phi_J}\right) \phi_i^{2}
	- \frac{\lambda}{4} \left( \phi_{i}^{2} \right)^{2}
	+ 36\alpha\left(\frac{\Box \Phi_J}{\Phi_J}\right)^2
	\right].
	\label{eq:NLSM_oN_quantum}
\end{align}
One can see that this expression coincides with 
the spin-0 sector of Eq.~\eqref{eq:Higgs_LSM_Jordan} as expected.
Namely, the field basis $\Phi_J$ convenient for the large-$N$ analysis corresponds to the Jordan frame.

Since the counter term~\eqref{eq:NLSM_higgs_ct} is a higher derivative term, 
it implies the existence of an additional degree of freedom.
To extract it, we introduce an auxiliary field $\sigma_J$ 
\begin{align}
	S  = \int \dd^4x \left\{
	- \frac{1}{2}\left(\partial \Phi_J\right)^2
	+ \frac{1}{2}\left(\partial \phi_i \right)^2 
	+ \frac{6\xi + 1}{2}\left(\frac{\Box \Phi_J}{\Phi_J}\right) \phi_i^{2}
	- \frac{\lambda}{4} \left( \phi_{i}^{2} \right)^{2}
	+ 36\alpha
	\left[
	\left(\frac{\Box \Phi_J}{\Phi_J}\right)^2
	- \left(\frac{\Box \Phi_J}{\Phi_J}+\frac{\Phi_J \sigma_J}{72\alpha}\right)^2
	\right]
	\right\}.
\end{align}
After shifting the fields as $\sigma_J = \sigma + (6\xi+1) \phi_i^{2} /2\Phi_J$
and  $\Phi = \Phi_J - \sigma$, we obtain the desired result
\begin{align}
	S = \int \dd^4x
	\left\{
	- \frac{1}{2}\left(\partial \Phi \right)^2
	+ \frac{1}{2}\left(\partial \sigma \right)^2
	+ \frac{1}{2}\left(\partial \phi_i \right)^2
	- \frac{\lambda}{4} \left( \phi_{i}^{2} \right)^{2}
	- \frac{1}{144\alpha}\left[ \frac{\Phi^2}{4} - \left(\sigma + \frac{\Phi}{2} \right)^2 - 
	\frac{6\xi+1}{2} \phi_i^{2} \right]^2
	\right\}.
	\label{eq:higgs-lsm}
\end{align}
Thus, the quantum correction in the large-$N$ limit, 
or the higher derivative term~\eqref{eq:NLSM_higgs_ct}, 
induces the $\sigma$-meson that linearizes the original NLSM \eqref{eq:nlsm-cl-higgs}.
It corresponds to the spin-0 sector of 
the Higgs-scalaron system~\eqref{eq:Higgs_LSM_cf},
and the additional degree of freedom corresponds to the scalaron.

\subsection{Remarks on the large-$N$ analysis}
\label{subsec:remarks_largeN}

Here are some remarks on our large-$N$ analysis.

\paragraph{Frame/gauge independence.}
Note that our large-$N$ analysis is frame and gauge independent.
As we have emphasized throughout this paper, 
the full result is guaranteed to be independent of the frame choice, 
and hence the result at each order in the large-$N$ expansion 
is also independent of this choice as one may vary $N$ arbitrarily.
The gauge independence of our results follows in the same way.

\paragraph{Cut-off scale in the large-$N$ limit.}
In this paper,
we have argued that the cut-off scale of Higgs inflation with the $R^2$ term is the Planck scale.
Strictly speaking, the cut-off scale is of order $M_P/\sqrt{N}$ if we take the large-$N$ limit.
This can be seen, \emph{e.g.}, from $d$-wave parts of scattering amplitudes 
or the RG running of the $R_{\mu\nu}R^{\mu\nu}$ term (see also Refs.~\cite{Han:2004wt,Dvali:2007hz}).
However, the typical scale of the $R^2$ term also scales in the same way,
and hence the fact that the spin-2 sector can be ignored in the large-$\xi$ limit is not affected.
For this reason, we have ignored this subtlety here.

\paragraph{Sub-leading terms in the large-$N$ expansion.}
In this section, we have relied on the large-$\xi$ and the large-$N$ limits.
Although the large-$\xi$ limit is expected to be good for $\xi = \mathcal{O}(10^4)$,
one may wonder how sub-leading terms in the large-$N$ expansion affect our understanding of Higgs inflation.
In the following, we suggest that
the LSM~\eqref{eq:higgs-lsm} provides a clue to answering this question.

As we have shown, the LSM~\eqref{eq:higgs-lsm}
describes the system to the leading order in the large-$N$ limit
if we ignore the spin-2 sector of gravity, which is valid in the large-$\xi$ limit.
Thus, sub-leading order terms in the large-$N$ limit can be obtained by 
computing quantum corrections of the LSM~\eqref{eq:higgs-lsm} below the Planck scale.
A significant feature of the LSM~\eqref{eq:higgs-lsm} is that
it possesses a flat target space and renormalizable interactions, 
and hence quantum corrections generate only a finite number of new operators.
Indeed, we can show that the LSM~\eqref{eq:higgs-lsm} with the Higgs mass term 
and the cosmological constant is renormalized at the one-loop level 
in the standard coupling expansion without any other new operators.
Hence we expect that, other than generating the Higgs mass term and the cosmological constant, 
sub-leading order terms do not affect our understanding of Higgs inflation.
In particular, we do not expect that operators such as $R^n$, with $n>2$, to be important below $M_P$,
since these higher-dimensional operators are not required to make the LSM~\eqref{eq:higgs-lsm} 
renormalizable, \emph{i.e.}, these are irrelevant operators.
In other words, we expect that the large-$\xi$ limit is sufficient for our understanding of Higgs inflation,
although we have relied on the large-$N$ limit to make our analysis simpler in this section.
Note that the $R^n$ terms with $n>2$ suppressed by $M_P$ are not expected to affect
the inflationary prediction of the Higgs-$R^2$ system for $\xi \sim \mathcal{O}(10^4)$
and  $\alpha \sim \xi^2$~\cite{Asaka:2015vza,Pi:2017gih,Cheong:2020rao}.

It is of course desirable to examine the above expectation by directly computing sub-leading order terms in the large-$N$ limit,
which we leave for future work.

\paragraph{Large-$N$ limit as a bottom up approach to UV theory.}
In this section, we have seen that the $\sigma$-meson or the scalaron emerges
and Higgs inflation is UV-completed to be the Higgs-scalaron system 
due to quantum corrections in the large-$N$ limit.
One might be surprised at our result since the UV-completion of a given theory is not unique,
and one cannot determine which UV-completion is chosen solely from an IR theory in general.
Here the large-$N$ limit does the trick.
Although there are many UV-completions of a given theory in general,
the large-$N$ limit naturally picks up one out of others.
In this sense, the large-$N$ limit provides us an interesting bottom up approach to UV completion.
In particular, if one has a non-renormalizable theory whose UV completion is not known,
the large-$N$ analysis will be a useful tool to find a possible UV completion.
For instance, it may be interesting to apply the large-$N$ analysis to, \emph{e.g.}, the Higgs effective field theory (EFT)~\cite{Feruglio:1992wf,Giudice:2007fh,Grinstein:2007iv,Azatov:2012bz,Alonso:2012px,Contino:2013kra,Buchalla:2013rka},
and try to extract a possible properties of the UV completion.
Note that the Higgs EFT can be formulated in terms of the target space
curvature~\cite{Alonso:2015fsp,Alonso:2016oah,Nagai:2019tgi,Helset:2020yio}, 
or equivalently regarded as a NLSM, 
and hence it is expected to be straightforward to apply the large-$N$ analysis to this theory.

\section{Summary and discussion}
\label{sec:summary}

\subsection{Summary}

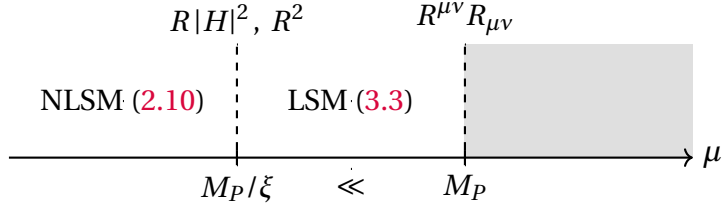
\begin{figure}[t]
	\centering
	\begin{tikzpicture} [thick]
	\fill[gray,opacity=.25] (6,0)--(9,0)--(9,1.5)--(6,1.5);
	\draw[->] (0,0)--(9,0) node[right] {\large $\mu$};
	\draw[-] (3,-0.1)--(3,0.1) node[below]{};
	\draw[dashed] (3,0.1)--(3,1.5) node[below]{};
	\draw[-] (3,-0.1)--(3,-0.1) node[below]{\large $M_P/\xi$};
	\draw[-] (3,1.5)--(3,1.5) node[above]{\large $R \abs{H}^{2},~R^2$};
	\draw[-] (4.5,-0.15)--(4.5,-0.15) node[below]{\large $\ll$};
	\draw[-] (6,-0.1)--(6,0.1) node[below]{};
	\draw[dashed] (6,0.1)--(6,1.5) node[below]{};
	\draw[-] (6,-0.1)--(6,-0.1) node[below]{\large $M_P$};
	\draw[-] (6,1.5)--(6,1.5) node[above]{\large $R^{\mu\nu}R_{\mu\nu}$};
	\draw[-] (1.5,0.75)--(1.5,0.75) node[]{\large NLSM~\eqref{eq:Higgs_NLSM_cf}};
	\draw[-] (4.5,0.75)--(4.5,0.75) node[]{\large LSM~\eqref{eq:Higgs_LSM_cf}};
	\end{tikzpicture}
	\caption{A schematic picture of the phase diagram of Higgs inflation 
	with $\xi \gg 1$, obtained with the help of the large-$N$ limit.
	In the low energy region, $\mu < M_P/\xi$, it is described by the NLSM~\eqref{eq:Higgs_NLSM_cf}
	with the Higgs and the conformal mode of the metric being the pions. 
	Once we go to the higher energy region, $M_P/\xi < \mu < M_P$,
	it is linearized as Eq.~\eqref{eq:Higgs_LSM_cf} with
	the scalaron playing the role of the $\sigma$-meson.
	In the even higher energy region $\mu > M_P$,
	other operators such as $R^{\mu\nu}R_{\mu\nu}$ come into play.
	One is probably required to fully take quantum gravity into account in this energy region, 
	which is beyond the scope of this paper.}
	\label{fig:summary}
\end{figure}

Higgs inflation introduces a nonminimal coupling $\xi$ 
between the Higgs $H$ and the Ricci scalar $R$ as $\xi R \abs{H}^2$.
The CMB normalization requires $\xi$ to be large, $\xi \gg 1$, 
unless the Higgs quartic coupling is tiny at the inflationary scale.
Consequences of this operator with a large value of $\xi$ have been studied in detail in the literature, 
including the tree-level unitarity violation at the energy scale  
$M_P/\xi \ll M_P$~\cite{Burgess:2009ea,Barbon:2009ya,Burgess:2010zq,Hertzberg:2010dc,Kehagias:2013mya,Burgess:2014lza}
and its implication during and after inflation~\cite{Bezrukov:2010jz,DeCross:2015uza, Ema:2016dny, Sfakianakis:2018lzf}.
Once we turn on quantum corrections, however,
for the large value of $\xi$, other operators are inevitably induced due to the RG running.
Among them, the most important one is the $R^2$ term, $\alpha R^2$, with its beta function given by Eq.~\eqref{eq:jordan-running}.
Due to this RG running, the natural mass scale of the scalaron that becomes dynamical due to the $R^2$ term
is $M_P/\sqrt{12\alpha} \sim M_P/\xi$, \emph{i.e.}, it becomes dynamical much below $M_P$. 
Since the scalaron can lift the cut-off scale to $M_P$~\cite{Ema:2017rqn, Gorbunov:2018llf},
this indicates that the tree-level unitarity violation mentioned above can be cured by 
quantum corrections~\cite{Aydemir:2012nz,Calmet:2013hia,Ema:2019fdd}.

In this paper, we have shown that Higgs inflation, the scalaron and its emergence
can be understood in the language of the nonlinear sigma model (NLSM)
in a frame independent way. 
In Sec.~\ref{sec:higgs_inf},
we have demonstrated that Higgs inflation can be written as a NLSM (see Eq.~\eqref{eq:Higgs_NLSM_cf}).
Our definition of the target space is frame independent
since we have included not only the scalar fields but also 
the conformal mode (or the determinant part) of the metric in our definition.
Thus, the Higgs fields and the conformal mode of the metric play the role of the \emph{pions}.
In the NLSM, we naturally expect an additional degree of freedom, the \emph{$\sigma$-meson}, 
that linearizes the NLSM. In Sec.~\ref{sec:scalaron}, we have shown that the scalaron plays 
the role of the $\sigma$-meson, and including this degree of freedom completely flattens the target space
and hence unitarizes the theory.
In Sec.~\ref{sec:large-N}, with the help of the large-$N$ limit, we have shown that 
the light $\sigma$-meson, or the scalaron, indeed appears due to quantum corrections.
Now described as a NLSM, our analysis in Sec.~\ref{sec:large-N} is clearly parallel to the large-$N$ analysis of, 
\emph{e.g.}, the $\mathrm{O}(N)$ NLSM~\cite{Bardeen:1976zh,Brezin:1976ap}, 
the $\mathrm{CP}^{N-1}$ model~\cite{Bando:1987br,DAdda:1978vbw,DAdda:1978dle,Witten:1978bc,Arefeva:1980ms,Fujiwara:1984mp},
the Nambu--Jona-Lasinio model~\cite{Nambu:1961tp,Nambu:1961fr}, and the Gross--Neveu model~\cite{Gross:1974jv}.

The phase diagram of Higgs inflation obtained with the help of the large-$N$ limit 
is summarized in Fig.~\ref{fig:summary}.
Higgs inflation is the NLSM~\eqref{eq:Higgs_NLSM_cf}
below the energy scale $M_P/\xi$, and it becomes the linear sigma model (LSM)~\eqref{eq:Higgs_LSM_cf} 
with the scalaron as the $\sigma$-meson above the scale $M_P/\xi$.

\subsection{Discussion}
\label{subsec:discussion}

We conclude this paper with some remarks that are not addressed in detail in the main text.

\paragraph{Heavy scalaron during inflation, fine tuning and perturbativity.}
Throughout this work, we have claimed that the natural mass scale of the scalaron is 
$M_P/\sqrt{12\alpha} \sim M_P/\xi$. 
Since $\alpha$ runs (according to Eq.~\eqref{eq:jordan-running} in the Jordan frame), 
its value depends on the boundary condition, or equivalently 
the choice of the scale $\Lambda$ at which $\alpha$ vanishes.\footnote{
	One should not confuse $\Lambda$ with the renormalization scale. 
	It is rather a model parameter as we explain just below.
} In this sense, we can think of $\Lambda$ instead of $\alpha$ as a model parameter, 
in the same way that we can think of $\Lambda_\mathrm{QCD}$ instead of the gauge coupling $g_3$ 
as a model parameter in QCD (this is called ``dimensional transmutation"~\cite{Coleman:1973jx}).
Therefore, one might choose $\Lambda$ 
such that the scalaron remains heavy during inflation
and the inflationary dynamics is described by Higgs inflation without the $R^2$ term.
Although possible, there are three subtleties one has to keep in mind in this scenario.
First, $\Lambda$ has to be tuned to be close to the inflationary scale 
for the scalaron to be heavy during inflation.
Hence this scenario requires tuning.
Second, due to the running of $\alpha$, it is impossible to keep the scalaron heavy for all energy scales for $\xi \gg 1$. 
Even if the scalaron is heavy during inflation, it becomes light after inflation and affects, \emph{e.g.}, reheating.
Finally, there is an issue related to perturbativity. As long as we rely on the large-$N$ limit, our analysis is valid for any value of $\alpha$.
If one computes quantities in the standard coupling expansion in the Higgs-scalaron system, however, perturbativity requires $\xi^2/4\alpha \lesssim 4\pi$,
and hence the small value of $\alpha$ implies that the system is in a strong coupling regime above $M_P/\xi$.

\paragraph{Unitarity during preheating.}
A very important consequence of the emergence of the $\sigma$-meson, or the scalaron, is 
that the unitarity cut-off scale of Higgs inflation can be lifted to the Planck scale 
(depending on the UV boundary condition of $\alpha$).
This feature is essential to follow the dynamics of Higgs inflation from inflation to reheating without ambiguity.
Although the energy scale of Higgs inflation at the classical level (\emph{i.e.} without the $R^2$ term) lies below the cut-off scale 
and does not necessarily lead to a problem \emph{during} inflation~\cite{Bezrukov:2010jz}, 
the story drastically changes \emph{after} inflation, during (p)reheating. 
After inflation, the Higgs field oscillates around the bottom of its potential.
When the Higgs field crosses zero, the strong curvature in the target space leads to violent production of longitudinal gauge bosons 
(or equivalently NG bosons), 
with momenta that seemingly violate the unitarity scale~\cite{DeCross:2015uza, Ema:2016dny, Sfakianakis:2018lzf}. 
Moreover, a naive estimate of the reheating temperature yields a value in the strong coupling regime. 
On the contrary, reheating with the $R^2$ term was studied in~\cite{He:2018mgb, Bezrukov:2019ylq}, 
where it was shown that the presence of the scalaron generally weakens particle production 
and unitarity is no longer violated by the produced particles.

\paragraph{RG flow of Higgs-scalaron system.}
Last but not least, we emphasize the power of the LSM~\eqref{eq:higgs-lsm}.
Although it contains only the scalar fields, 
it is expected to correctly reproduce quantum effects of the theory up to the Planck scale or a Landau pole.
For instance, we can compute the RGEs of the dimensionless parameters and the ratios of the dimensionful parameters at the one-loop level
within the LSM~\eqref{eq:higgs-lsm},
and we can show that they agree with the scalar part of the full computation within quadratic gravity in 
Refs.~\cite{Salvio:2014soa,Salvio:2017qkx,Salvio:2018crh}.\footnote{
	Indeed, a scalar field model is discussed in~\cite{Salvio:2017qkx} that correctly reproduces the running of $\xi$ and $\alpha$ 
	(or $f_0$ in their language), which is similar to our LSM~\eqref{eq:higgs-lsm}.
} Note that the LSM~\eqref{eq:higgs-lsm} greatly simplifies the computation
since it does not contain any tensor modes.
In particular, we can see that the Higgs mass term and the cosmological constant are radiatively generated
even if they are absent at a specific energy scale.
This is because the scalaron introduces an additional mass scale, 
and it can be understood as a specific form of the hierarchy problem.
See Ref.~\cite{Ema:2020evi} for more details on this point.

\section*{Acknowledgement}
The authors would like to thank Valerie Domcke and Ben Mares for helpful discussions and comments.
This work was funded by the Deutsche Forschungsgemeinschaft under Germany’s Excellence
Strategy - EXC 2121 “Quantum Universe” - 390833306.
This work is also supported by the ERC Starting Grant `NewAve’ (638528).
The Feynman diagrams in this paper are drawn by \texttt{TikZ-Feynman}~\cite{Ellis:2016jkw}. 

\appendix

\section{Conventions}
\label{app:convention}

Here we summarize our conventions.
In this paper we work with the mostly-minus convention for the spacetime metric.
In particular, the flat spacetime metric is given by
\begin{align}
	\eta_{\mu\nu} = \mathrm{diag}\left(+1, -1, -1, -1\right).
\end{align}
We define the Christoffel symbol as
\begin{align}
	{\Gamma^{\mu}}_{\nu\rho} = \frac{1}{2}g^{\mu\alpha}
	\left(\partial_\nu g_{\rho \alpha} + \partial_\rho g_{\nu \alpha} - \partial_\alpha g_{\nu\rho}\right),
\end{align}
the Ricci tensor as
\begin{align}
	R_{\mu\nu} = \partial_\mu {\Gamma^{\alpha}}_{\alpha \nu} 
	- \partial_{\alpha} {\Gamma^{\alpha}}_{\mu\nu}
	+ {\Gamma^{\alpha}}_{\beta \nu} {\Gamma^{\beta}}_{\alpha\mu}
	- {\Gamma^{\alpha}}_{\mu\nu}{\Gamma^{\beta}}_{\alpha\beta},
\end{align}
and the Ricci scalar as
\begin{align}
	R = g^{\mu\nu}R_{\mu\nu}.
\end{align}
This fixes the sign convention for the Ricci scalar. In particular, 
the Ricci scalar transforms under the Weyl transformation $g_{\mu\nu} \rightarrow \Omega^{-2}g_{\mu\nu}$ as
\begin{align}
	R \rightarrow \Omega^2 \left[R + \frac{3}{2}g^{\mu\nu}\partial_{\mu} \ln \Omega^2 \partial_{\nu} \ln \Omega^2
	- 3 \Box \ln \Omega^2 \right].
\end{align}
The conformal coupling corresponds to 
$\xi = -1/6$ with this convention.

\section{Gauge fixing and residual gauge symmetry}
\label{sec:gauge}
In this appendix, we discuss gauge fixing conditions and residual gauge symmetries.
In Sec.~\ref{sec:large-N}, we have focused on the conformal mode of the metric $\Phi$.
Here we show a gauge fixing condition that corresponds to this treatment; 
see Eqs.~\eqref{eq:gauge_gravity}.
We also confirm that the ghost-like field $\Phi$ is indeed harmless due to the residual gauge symmetry; 
see Eqs.~\eqref{eq:residual_gravity},~\eqref{eq:residual_gravity_component} and~\eqref{eq:residual_gravity_varphi}.

\subsection{U(1) gauge theory}
\label{subsec:U(1)}
As a warm-up, we consider the U(1) gauge theory in this subsection.
The discussion is quite parallel to the gravity case, and hence it is useful to understand this simpler case first.
We consider the U(1) gauge field $A_\mu$ that transforms under a U(1) gauge transformation as
\begin{align}
	A_{\mu} \rightarrow A_{\mu} + \partial_{\mu}\theta.
	\label{eq:gauge_U(1)}
\end{align}
We may impose the Lorenz gauge condition
\begin{align}
	\partial^{\mu} A_{\mu} = 0,
	\label{eq:gauge_Lorenz}
\end{align}
to kill one degree of freedom. It still has a residual gauge symmetry. Indeed, one can perform 
the transformation~\eqref{eq:gauge_U(1)} without affecting Eq.~\eqref{eq:gauge_Lorenz} provided $\theta$ satisfies
\begin{align}
	\Box \theta = 0,
	\label{eq:residual_Lorenz}
\end{align}
that makes another degree of freedom unphysical.
As a result, there are two physical modes in $A_{\mu}$ that correspond to the two polarizations of the photon.
Note that the Lagrangian for $A_{\mu}$ is given after imposing Eq.~\eqref{eq:gauge_Lorenz} by
\begin{align}
	\mathcal{L} = -\frac{1}{4}F_{\mu\nu}F^{\mu\nu}
	= -\frac{1}{2}\eta^{\mu\nu}\partial_{\rho}A_{\mu} \partial^{\rho}A_{\nu},
\end{align}
and hence the time-like component $A_0$ is ghost-like, as in the case of $\Phi$ in the main text.
It is harmless due to the residual gauge symmetry~\eqref{eq:residual_Lorenz}. The Gupta-Beuler condition guarantees that all physical states are healthy \cite{1950PPSA...63..681G, Bleuler1950}.

In order to take a closer look at the degrees of freedom in $A_{\mu}$ 
killed by Eqs.~\eqref{eq:gauge_Lorenz} and~\eqref{eq:residual_Lorenz},
we decompose $A_{\mu}$ as
\begin{align}
	A_{\mu} = A_{\mu}^{\perp} + \partial_{\mu} A,
	\label{eq:decomposition_U(1)}
\end{align}
where $A_{\mu}^{\perp}$ satisfies
\begin{align}
	\partial^{\mu}A_{\mu}^{\perp} = 0.
\end{align}
It is important to notice that there is an ambiguity in the decomposition~\eqref{eq:decomposition_U(1)};
we can shift $A_{\mu}^{\perp}$ and $A$ as
\begin{align}
	A_{\mu}^{\perp} \rightarrow A_{\mu}^{\perp} + \partial_{\mu}B,
	~~
	A \rightarrow A - B,
	~~~\mathrm{with}~~~
	\Box B = 0,
	\label{eq:shiftA}
\end{align}
without spoiling the transverse property of $A_{\mu}^{\perp}$.
Due to this ambiguity, it is enough to require
\begin{align}
	\Box A = 0,
	\label{eq:gauge_Lorenz_component}
\end{align}
to kill the degree of freedom corresponding to $A$.
Indeed, the Lorenz gauge condition~\eqref{eq:gauge_Lorenz} requires Eq.~\eqref{eq:gauge_Lorenz_component},
and hence kills $A$.
The residual gauge symmetry~\eqref{eq:residual_Lorenz} kills an additional degree of freedom in $A_{\mu}^{\perp}$ 
that is ghost-like without affecting the condition~\eqref{eq:gauge_Lorenz_component}.

\subsection{Gravity}
\label{subsec:gravity}
Now we consider gravity. We may expand the metric as
\begin{align}
	g_{\mu\nu} = e^{2\varphi}\left(\eta_{\mu\nu} + h_{\mu\nu}\right),
\end{align}
and treat $h_{\mu\nu}$ as a perturbation as we have done in the main text.
Note that we do \textit{not} treat the conformal mode as a perturbation.
Under the general coordinate transformation,
\begin{align}
	x^{\mu} \rightarrow x^{\mu} - \xi^{\mu},
\end{align}
the modes transform at the first order in $h_{\mu\nu}$ and $\xi^\mu$ as
\begin{align}
	\varphi &\rightarrow \varphi + \frac{1}{4}\partial_\alpha \xi^\alpha + \xi^\alpha \partial_\alpha \varphi, 
	\label{eq:transf_gravity_scalar} \\
	h_{\mu\nu} &\rightarrow h_{\mu\nu} + \partial_{\mu}\xi_{\nu} + \partial_{\nu} \xi_{\mu} 
	- \frac{1}{2}\eta_{\mu\nu} \partial_{\alpha}\xi^{\alpha},
	\label{eq:transf_gravity_tensor}
\end{align}
where the indices are raised and lowered by the flat spacetime metric $\eta_{\mu\nu}$ here and hereafter
in this subsection.
We may impose a gauge fixing condition
\begin{align}
	\partial^{\mu}h_{\mu\nu} = 0,
	\label{eq:gauge_gravity}
\end{align}
which is slightly different from the standard de~Donder gauge.
It kills four degrees of freedom, which leaves six degrees of freedom in $h_{\mu\nu}$ and $\varphi$.
Among them, four are killed by a residual gauge symmetry as in the U(1) case.
Indeed, we can still perform the transformations~\eqref{eq:transf_gravity_scalar} and~\eqref{eq:transf_gravity_tensor} 
without affecting Eq.~\eqref{eq:gauge_gravity} if $\xi_{\mu}$ satisfies
\begin{align}
	\Box \xi_\mu + \frac{1}{2}\partial_\mu \partial^{\nu} \xi_{\nu} = 0.
	\label{eq:residual_gravity}
\end{align}
Thus, there are two physical modes that correspond to the two polarizations of the graviton.

We now take a closer look at the degrees of freedom killed by Eqs.~\eqref{eq:gauge_gravity} and~\eqref{eq:residual_gravity}. 
We decompose the traceless part of the metric as
\begin{align}
	h_{\mu\nu} = h_{\mu\nu}^{\perp} + \partial_{\mu} h_{\nu}^{\perp} + \partial_{\nu} h_{\mu}^{\perp}
	+ \left(\partial_{\mu}\partial_{\nu} - \frac{1}{4}\eta_{\mu\nu}\Box\right)\psi,
	\label{eq:decomposition_gravity}
\end{align}
where $h^{\perp}_{\mu\nu}$ and $h^{\perp}_{\mu}$ satisfy
\begin{align}
	{h^{\perp \mu}}_{\mu} = \partial^{\mu}h_{\mu\nu}^{\perp} = 0,
	~~
	\partial^{\mu}h^{\perp}_{\mu} = 0.
\end{align}
As in the U(1) case, there are ambiguities in this decomposition. 
Indeed, $\psi$ can be absorbed into $h_{\mu\nu}^{\perp}$ and $h_{\mu}^{\perp}$ if it satisfies
\begin{align}
	 \left(\partial_{\mu}\partial_{\nu} - \frac{1}{4}\eta_{\mu\nu}\Box\right)\psi  
	 = f^{\perp}_{\mu\nu} + \partial_{\mu}f^{\perp}_{\nu} + \partial_{\nu}f^{\perp}_{\mu},
\end{align}
where $f^{\perp}_{\mu\nu}$ and $f^{\perp}_{\mu}$ satisfy the same properties as 
$h^{\perp}_{\mu\nu}$ and $h^{\perp}_{\mu}$, respectively.
By acting with $\partial^{\mu}\partial^{\nu}$, we see that it is enough to require
\begin{align}
	\Box^2 \psi = 0,
	\label{eq:gauge_gravity_component}
\end{align}
to kill the degree of freedom associated with $\psi$.
The gauge fixing condition~\eqref{eq:gauge_gravity} reduces to this condition after acting with $\partial^\mu$,
and hence kills $\psi$.
Thus, it is indeed Eq.~\eqref{eq:gauge_gravity} that we have imposed in the main text
since we have focused only on $\varphi$ and eliminated $\psi$ there.
It is also easy to see that the residual gauge symmetry makes 
the conformal mode $\varphi$ (or equivalently $\Phi$) unphysical.
If we write $\xi_{\mu} = \xi_{\mu}^{\perp} + \partial_{\mu} \xi$ with $\partial^{\mu}\xi_{\mu}^{\perp} = 0$, 
the gauge fixing condition~\eqref{eq:gauge_gravity_component} is intact
as long as $\xi$ satisfies
\begin{align}
	\Box^2 \xi = 0.
	\label{eq:residual_gravity_component}
\end{align}
Since $\varphi$ transforms as
\begin{align}
	\varphi \rightarrow \varphi + \frac{1}{4} \Box \xi + \eta^{\mu\nu}\partial_\mu \xi \partial_\nu \varphi,
	\label{eq:residual_gravity_varphi}
\end{align}
it becomes unphysical due to this residual gauge symmetry.
Note that Eq.~\eqref{eq:residual_gravity_component} is indeed one of the residual gauge symmetries of~\eqref{eq:residual_gravity},
since the latter is equivalent to
\begin{align}
	\Box \xi_{\mu}^{\perp} + \frac{3}{2}\partial_{\mu}\Box \xi = 0,
\end{align}
and hence we obtain Eq.~\eqref{eq:residual_gravity_component} by acting with $\partial^{\mu}$.

Before closing this appendix, we mention another residual gauge symmetry in the vector-tensor sector for completeness.
First of all, there is another ambiguity in the decomposition~\eqref{eq:decomposition_gravity};
$h_{\mu}^{\perp}$ can be absorbed into $h_{\mu\nu}^{\perp}$ if it satisfies
\begin{align}
	\partial_{\mu}h_{\nu}^{\perp} + \partial_{\nu}h_{\mu}^{\perp} = f_{\mu\nu}^{\perp},
\end{align}
where $f_{\mu\nu}^{\perp}$ satisfies the same properties as $h_{\mu\nu}^{\perp}$.
As a result, it is enough to require
\begin{align}
	\Box h_{\mu}^{\perp} = 0,
	\label{eq:gauge_gravity_vector}
\end{align}
to kill the degree of freedom associated with $h_{\mu}^{\perp}$,
which can be derived from Eq.~\eqref{eq:gauge_gravity}.
The transformation~\eqref{eq:transf_gravity_tensor} keeps Eq.~\eqref{eq:gauge_gravity_vector} intact as long as 
$\xi_{\mu} = \xi_{\mu}^{\perp}$ and $\xi_{\mu}^{\perp}$ satisfies
\begin{align}
	\Box \xi_{\mu}^{\perp} = 0.
\end{align}
It is the residual gauge symmetry that kills unphysical modes in $h_{\mu\nu}^{\perp}$. In the case of a non-Abelian theory including gravity, 
the Kugo-Ojima condition \cite{Kugo:1979gm} guarantees that the physical states contain no ghosts. 

\section{Large-$N$ analysis of generalized model}
\label{app:general}

In this section, we generalize the large-$N$ analysis in Sec.~\ref{subsec:higgs_inf}.
Let us start with
\begin{align}
	S = \int \dd^{4} x \left\{
		- \frac{1}{2} \left( \partial \Phi \right)^{2} + \frac{1}{2} \left( \partial \phi_{i} \right)^{2} + \frac{1}{2} \left[ \partial \left( \sqrt{a^{2} \Phi^{2} - b \phi_{i}^{2}} - c \Phi \right) \right]^{2}
	\right\},
	\label{eq:NLSM_abc}
\end{align}
which is a slight generalization of the spin-0 sector of Eq.~\eqref{eq:Higgs_NLSM_cf}.
Its target space is
\begin{align}
	b \phi_{i}^{2} + \left(h + c \Phi \right)^{2} = a^{2} \Phi^{2}
	\quad \text{in} \quad \left(\Phi, \phi_{i}, h \right) \in \mathbb{R}^{(1, N+1)},
\end{align}
which is an $N+1$-dimensional hypersurface in $\mathbb{R}^{(1, N+1)}$.
Higgs inflation corresponds to $a = c = 1/2$ and $b = (6 \xi + 1) / 2$.
We can show that this choice of parameters is special as it allows us to have successful inflation as discussed in App.~\ref{sec:so11}.\footnote{
	As we see in App.~\ref{sec:so11}, there are redundancies in the parameters $a$, $b$, and $c$ from transformation of fields (coordinate transformation of the target space). Here we mean ``special'' up to these redundancies.
}
In this appendix, the contraction of the Lorentz indices is always taken by $\eta_{\mu\nu}$
unless otherwise specified.

In order to perform the large-$N$ analysis, the field basis in Eq.~\eqref{eq:NLSM_abc} is not convenient because it involves a square root.
We thus perform a field redefinition
so that the large-$N$ analysis becomes more transparent:
\begin{align}
	 \Phi_J = \sqrt{a^{2} \Phi - b \phi_{i}^{2}} + a \Phi,
\end{align}
which implies
\begin{align}
	\frac{1}{2}\left(\Phi_J - \frac{b \phi_{i}^{2}}{\Phi_J}\right) 
	&= \sqrt{a^2 \Phi^2 - b\phi_{i}^{2}} ,
	\qquad \frac{1}{2}\left(\Phi_J + \frac{b \phi_{i}^{2}}{\Phi_J}\right) 
	= a \Phi.
\end{align}
We put the subscript $J$ since it indeed corresponds to $\Phi_J$ in the main text in the case of Higgs inflation.
By using $\Phi_J$, one may write down Eq.~\eqref{eq:NLSM_abc} as follows:
\begin{align}
	S = \int \dd^{4} x \left\{
		\frac{1}{2} \left( \partial \phi_{i} \right)^{2}
		- \frac{1}{8 a^{2}} \left[ \partial \left( \Phi_J + \frac{b \phi_{i}^{2}}{\Phi_J} \right) \right]^{2}
		+ \frac{1}{8} \left[ \partial \left( \left( 1 - \frac{c}{a} \right) \Phi_J  
		- \left( 1 + \frac{c}{a} \right) \frac{b \phi_{i}^{2}}{\Phi_J} \right) \right]^{2}
	\right\},
	\label{eq:NLSM_abc2}
\end{align}
which now contains only a finite number of $\phi_{i}^{2}$-interactions.\footnote{
	It does not matter that $\tilde{\Phi}$ appears in the denominator
	since we have to care only about $\phi_{i}$ in the large-$N$ limit.
}
It contains two types of interactions
\begin{align}
	\left(\frac{\Box {\Phi_J}}{\Phi_J}\right) \phi_i^{2},
	\qquad
	\left[\partial \left(\frac{\phi_i^{2}}{\Phi_J}\right)\right]^2.
\end{align}
We can find divergences and counter terms by taking both of these interactions into account in the large-$N$ limit.
Instead, here we introduce two vector auxiliary fields, $\rho_\mu$ and $A_\mu$, to reduce the number of relevant interactions further.
With these fields, we rewrite the action~\eqref{eq:NLSM_abc2} as
\begin{align}
	S  = \int \dd^4x \left\{
	\frac{1}{2}\left(\partial \phi_{i} \right)^2 
	+ \frac{b}{2}\rho^\mu \partial_\mu \left(\frac{\phi_i^{2}}{{\Phi}_J}\right)
	+ \frac{1}{2}\partial_\mu {\Phi}_J \left(\rho^\mu - 2 A^\mu\right)
	+ \frac{1}{2}\left[ a^2 \left(\rho_\mu - \left(1 + \frac{c}{a}\right) A_\mu\right)^2
	- A_\mu A^\mu \right]
	\right\}.
	\label{eq:NLSM_abc_classical}
\end{align}
The interaction of $\phi_i$ is now contained entirely in the term
\begin{align}
	\mathcal{L}_{\mathrm{int}} = \frac{b}{2}\rho^\mu\partial_{\mu}\left(\frac{\phi_i^{2}}{{\Phi}_J}\right)
	= - \frac{b \phi_i^{2}}{2 {\Phi}_J} \partial_\mu \rho^\mu + \left(\text{total derivative}\right),
\end{align}
and hence the computation below is greatly simplified. 
We emphasize here that it is merely for convenience,
and the final result should not change even if 
we do not introduce the vector auxiliary fields.

We now study quantum corrections to the action~\eqref{eq:NLSM_abc_classical}.
The new divergence only appears in the two-point function of the operator 
$\partial_\mu \rho^\mu/ \Phi_J$ in the large-$N$ limit,
which at the one-loop level is diagrammatically given by
\begin{align}
	\begin{tikzpicture}[baseline=(c)]
	\begin{feynman}[inline]
		\vertex (c);
		\vertex [right = of c] (d);
		\vertex [right = of d] (e);
		\vertex [right = of e] (f);
		\diagram*{
		(c) -- [photon] (d),
		(e) -- [scalar, half left] (d) -- [scalar, half left] (e),
		(e) -- [photon] (f)
		};
	\end{feynman}
	\end{tikzpicture}
\end{align}
where the wavy line indicates the operator $\partial_\mu \rho^\mu/ \Phi_J$
and the dotted line denotes the scalar fields $\phi_i$.
We have to introduce the following operator as a counter term:
\begin{align}
	\mathcal{L}_\mathrm{c.t.} = 9\alpha\left(\frac{\partial_\mu \rho^\mu}{\Phi_J}\right)^2.
	\label{eq:NLSM_abc_ct}
\end{align}
Note that the leading order terms at the higher loop level, which are diagrammatically given by, 
\begin{align}
	\begin{tikzpicture}[baseline=(c)]
	\begin{feynman}[inline]
		\vertex (c);
		\vertex [right = of c] (d);
		\vertex [right = of d] (e);
		\vertex [right = of e] (f);
		\vertex [right = of f] (g);
		\vertex [right = of g] (h);
		\diagram*{
		(c) -- [photon] (d),
		(e) -- [scalar, half left] (d) -- [scalar, half left] (e),
		(e) -- [photon] (f),
		(g) -- [scalar, half left] (f) -- [scalar, half left] (g),
		(g) -- [photon] (h)
		};
	\end{feynman}
	\end{tikzpicture}
	+
	\begin{tikzpicture}[baseline=(c)]
	\begin{feynman}[inline]
		\vertex (c);
		\vertex [right = of c] (d);
		\vertex [right = of d] (e);
		\vertex [right = of e] (f);
		\vertex [right = of f] (g);
		\vertex [right = of g] (h);
		\vertex [right = of h] (i);
		\vertex [right = of i] (j);
		\diagram*{
		(c) -- [photon] (d),
		(e) -- [scalar, half left] (d) -- [scalar, half left] (e),
		(e) -- [photon] (f),
		(g) -- [scalar, half left] (f) -- [scalar, half left] (g),
		(g) -- [photon] (h),
		(i) -- [scalar, half left] (h) -- [scalar, half left] (i),
		(i) -- [photon] (j)
		};
	\end{feynman}
	\end{tikzpicture}
	+
	\cdots
\end{align}
are also renormalized by the same term~\eqref{eq:NLSM_abc_ct} as in Sec.~\ref{sec:large-N}. 
After the renormalization, the coupling $\alpha$ runs according to the beta function as
\begin{align}
	\frac{\dd \alpha}{\dd \ln \mu} = -\frac{N}{288\pi^2}b^2,
	\label{eq:alphaRG_abc}
\end{align}
in the large-$N$ limit.

By including the term generated by the quantum correction \eqref{eq:NLSM_abc_ct}, the action is now given by
\begin{align}
	S = \int \dd^4x &\Bigg\{
	\frac{1}{2}\left(\partial \phi_i \right)^2 
	+ \frac{b}{2}\rho^\mu \partial_\mu \left(\frac{\phi_i^{2}}{{\Phi_J}}\right)
	+ 9\alpha \left(\frac{\partial_\mu \rho^\mu}{{\Phi_J}}\right)^2
	\nonumber \\ &
	+ \frac{1}{2}\partial_\mu {\Phi_J} \left(\rho^\mu - 2 A^\mu\right)
	+ \frac{1}{2}\left[ a^2 \left(\rho_\mu - \left(1 + \frac{c}{a}\right) A_\mu\right)^2
	- A_\mu A^\mu \right]
	\Bigg\}.
	\label{eq:NLSM_abc_quantum}
\end{align}
Now $\rho_\mu$ obtained a kinetic term due to quantum corrections, which implies the appearance of a new degree of freedom.
In order to extract it in a simpler form, we introduce a scalar auxiliary field ${\sigma_J}$ as
\begin{align}
	S  = \int \dd^4x &\Bigg\{
	\frac{1}{2}\left(\partial \phi_i \right)^2 
	+ \frac{b}{2}\rho^\mu \partial_\mu \left(\frac{\phi_i^{2}}{{\Phi_J}}\right)
	+ 9\alpha \Bigg[ \left(\frac{\partial_\mu \rho^\mu}{{\Phi_J}}\right)^2
	- \left(\frac{\partial_\mu \rho^\mu}{{\Phi_J}}
	+\frac{1}{36\alpha} \left({\Phi_J} {\sigma_J} - b\phi_i^{2} \right)\right)^2
	\Bigg]
	\nonumber \\ &
	+ \frac{1}{2}\partial_\mu {\Phi_J} \left(\rho^\mu - 2 A^\mu\right)
	+ \frac{1}{2}\Bigg[ a^2 \left(\rho_\mu - \left(1 + \frac{c}{a}\right) A_\mu\right)^2
	- A_\mu A^\mu \Bigg]
	\Bigg\}.
\end{align}
Performing integration by parts and shifting $\rho_\mu$ 
as $\rho_\mu \rightarrow \rho_\mu + (1+c/a)A_\mu$,
we obtain
\begin{align}
	S = \int \dd^4x &\Bigg\{
	\frac{1}{2}\left(\partial \phi_i \right)^2 
	+ \frac{1}{2}A^\mu \partial_\mu\left[
	\left(1+\frac{c}{a}\right){\sigma}_J - \left(1-\frac{c}{a}\right){\Phi}_J
	\right]
	- \frac{1}{144\alpha}\left({\Phi}_J {\sigma}_J - b \phi_i^{2} \right)^2
	\nonumber \\ &
	+ \frac{1}{2}\rho^\mu \partial_\mu \left({\Phi}_J + {\sigma}_J\right)
	+ \frac{1}{2}\left(a^2 \rho_\mu \rho^\mu - A_\mu A^\mu \right)
	\Bigg\}.
	\label{eq:NLSM_abc_quantum_chi}
\end{align}
At this stage, the derivatives are not acting on $\rho_\mu$ and $A_\mu$ any more,
and hence we can integrate them out without introducing non-local terms.
By further redefining the fields as
\begin{align}
	\Phi &\equiv \frac{1}{2a}\left({\Phi}_J + {\sigma}_J \right), \qquad
	\sigma \equiv \frac{1}{2}\left[\left(1-\frac{c}{a}\right){\Phi}_J - \left(1+\frac{c}{a}\right){\sigma}_J\right],
\end{align}
we finally obtain
\begin{align}
	S  = \int \dd^4 x 
	\left\{
	-\frac{1}{2}\left(\partial \Phi \right)^2 + \frac{1}{2}\left(\partial \phi_i \right)^2
	+ \frac{1}{2}\left(\partial \sigma\right)^2
	- \frac{1}{144\alpha}\left[a^2 \Phi^2 - \left(\sigma + c \Phi\right)^2 - b \phi_i^{2} \right]^2
	\right\}.
	\label{eq:LSM_abc}
\end{align}
Thus, the additional degree of freedom is indeed the $\sigma$-meson that UV-completes the original NLSM as a LSM for the general case with arbitrary $a, b$ and $c$.

The corresponding action in the conformal frame is obtained by identifying $\Phi = \sqrt{6} M_{P} e^{\varphi}$ and recalling $g_{\mu\nu} = e^{2 \varphi} \eta_{\mu\nu}$:
\begin{align}
	S = \int \dd^4x \sqrt{-g}
	&\Bigg\{
	\frac{R}{12}\left(6 M_{P}^{2} - \sigma_C^2 - \phi_{Ci}^2\right)
	+ \frac{1}{2}g^{\mu\nu} \partial_\mu \sigma_C \partial_\nu \sigma_C
	+ \frac{1}{2}g^{\mu\nu} \partial_\mu \phi_{Ci} \partial_\nu \phi_{Ci}
	\nonumber \\ &
	- \frac{1}{144\alpha}\left[6 a^2 M_{P}^{2}  - \left(\sigma_C + c \sqrt{6}M_{P} \right)^2 - b \phi_{Ci}^2 \right]^2
	\Bigg\},
	\label{eq:lsm-abc-grav}
\end{align}
where the scalar fields are also rescaled as $\sigma_C = e^{-\varphi} \sigma$ and $\phi_{Ci} = e^{-\varphi} \phi_i$.
We can verify that the running of the mass term within the UV theory~\eqref{eq:LSM_abc}
and the RG running of $\alpha$~\eqref{eq:alphaRG_abc} computed within the IR theory agrees with each other in the large-$N$ limit.

\section{$\mathrm{O}(1,1)$ transformation and flat potential}
\label{sec:so11}

In this appendix, starting from the generalized model introduced in App.~\ref{app:general}, we discuss the condition to have a flat potential suitable for inflation.
Our starting point is
\begin{align}
	S  = \int \dd^4 x 
	\left[
	-\frac{1}{2}\left(\partial \Phi \right)^2 + \frac{1}{2}\left(\partial \phi_i \right)^2
	+ \frac{1}{2}\left(\partial \sigma\right)^2 - V (\Phi, \sigma, \phi_{i}^{2})
	\right],
	\label{eq:lsm-general}
\end{align}
where
\begin{align}
	V (\Phi, \sigma, \phi_{i}^{2}) \equiv \frac{\lambda}{4} \left( \phi_{i}^{2} \right)^{2}
	+ \frac{1}{144\alpha}\left[a^2 \Phi^2 - \left(\sigma + c \Phi\right)^2 - b \phi_i^{2} \right]^2.
	\label{eq:potential}
\end{align}
Note that we can take $a, c \geq 0$ without loss of generality.
Higgs inflation with the $\sigma$-meson or the scalaron corresponds to a particular set of parameters $a = c = 1/2$ and $b = (6 \xi + 1)/2$.
The main purpose of this appendix is to clarify why this choice of parameters yields a flat potential suitable for inflation and how special this choice is.

\subsection{Flat potential}
We first clarify the condition to have a potential which approaches asymptotically to a constant value in the Einstein frame.
One may go to the Einstein frame by identifying: $\Phi^{2} - \sigma^{2} - \phi_{i}^{2} = 6 M_{P}^{2} e^{2 \varphi}$.
In the following discussion, gravity is irrelevant and hence we may take $\varphi = 0$:
\begin{align}
	\Phi^{2} - \sigma^{2} - \phi_{i}^{2} = 6 M_{P}^{2}. 
	\label{eq:einstein-fix}
\end{align}
Let $(\Phi_{\theta},\sigma_{\theta},\phi_{i,\theta})$ be a one-dimensional trajectory, \emph{i.e.}, $\mathbb{R} \to \mathbb{R}^{1,N+1}$; $\theta \mapsto (\Phi_{\theta},\sigma_{\theta},\phi_{i,\theta})$, fulfilling Eq.~\eqref{eq:einstein-fix}.
We have a flat direction in the potential in the Einstein frame if we find a trajectory $\theta$ on which the potential $V (\Phi_{\theta}, \sigma_{\theta}, \phi_{i, \theta}^{2})$ approaches asymptotically to a constant or does not change at all.

Since the potential should be finite along this trajectory, $\phi_{i, \theta}^{2}$ is bounded from above because of the $\lambda (\phi_{i}^{2})^{2}$ term.
A trivial example fulfilling these requirements is the NG boson directions of the Higgs. There $\phi_{i, \theta}^{2}$, $\sigma_{\theta}$, and $\Phi_{\theta}$ are fixed to be constants.
What we are interested in here is a less trivial trajectory. 
Namely, $\Phi_{\theta}$ and $\sigma_{\theta}$ can be taken to infinity because of a non-trivial cancellation among them in the second term in Eq.~\eqref{eq:potential},
while $\phi_{i,\theta}^{2}$ is bounded from above.
In order to have this trajectory, one should find a trajectory of $V \to \text{const.}$ for $\Phi_{\theta}, \sigma_{\theta} \to \infty$ even under
\begin{align}
	\Phi_{\theta}^{2} - \sigma_{\theta}^{2} = \Lambda^{2}, \quad
	\phi_{i,\theta}^{2} = \Lambda^{2} - 6 M_{P}^{2} > 0,
	\label{eq:cond-o11}
\end{align}
with $\Lambda$ being a constant.
In the following, we discuss the impact of this condition on $a,b$, and $c$.

The trajectory fulfilling Eq.~\eqref{eq:cond-o11} can be expressed by a single parameter $\theta$ as
\begin{align}
	\Phi = \Lambda \cosh \theta, \quad \sigma = \Lambda \sinh \theta.
	\label{eq:trj}
\end{align}
Inserting this expression into the potential, one obtains the following form for the second term in Eq.~\eqref{eq:potential}:
\begin{align}
	\frac{1}{144\alpha} \left\{ \Lambda^{2} \left[ \sinh \theta + (a + c) \cosh \theta \right]  \left[ \sinh \theta - (a - c) \cosh \theta \right] + b (\Lambda^{2} - 6 M_{P}^{2}) \right\}^{2}.
\end{align}
In order for the potential to approach asymptotically to a constant value for $\theta \to \pm \infty$, the following condition should be fulfilled:
\begin{align}
	a + c = 1 \,\,\, \lor \,\,\, a - c = 1 \,\,\, \lor \,\,\, a - c = -1 \,\,\, \lor \,\,\, a + c = - 1.
\end{align}
As mentioned earlier, we can take $a, c \geq 0$ without loss of generality, and hence we focus on the first three branches in the following.
In the second and third branches, our vacuum in the current Universe $\phi_{i}^{2} = 0$ (which is also a potential minimum) is located at $\abs{\theta} = \infty$ for $a, c \geq 0$, \emph{i.e.}, a run-away potential.
Similarly, one readily finds a run-away potential for the first branch at $a = 0$.
For the first branch with $a = 1$ and $c = 0$, on the other hand, one ends up with an exactly massless mode which is completely decoupled from the Higgs field $\phi_{i}$.
These cases might not be interesting in the context of Higgs inflation.

Therefore we arrive at the case with
\begin{align}
	a + c = 1, \quad a > 0, \quad c > 0.
	\label{eq:prm-inf}
\end{align}
As we show below, this case is equivalent to Higgs inflation, $a = c = 1/2$ and $b = (6 \xi + 1)/2$, after appropriately redefining the parameters.

Here we comment on the physical meaning of Eq.~\eqref{eq:prm-inf}.
In this case, the potential becomes flat in the large $\theta$ direction
and the O$(1,1)$ symmetry between $\Phi$ and $\sigma$ gets restored.
In the Jordan frame language, it corresponds to classical scale invariance.
During inflation, the $R^2$ term and the nonminimal coupling become more important than
the Einstein-Hilbert term.
This means that the Planck scale can be ignored and hence the theory has classical scale invariance.

\subsection{Redundancy in parameters and O$(1,1)$ transformation}

In this section, we point out redundancies in the parameters $a$, $b$, and $c$.
To this end, the following O$(1,1)$ transformation plays a central role:
\begin{align}
	\begin{pmatrix}
	\Phi' \\ \sigma' 
	\end{pmatrix}
	=
	\begin{pmatrix}
	\cosh \theta & - \sinh \theta \\ 
	- \sinh \theta & \cosh \theta
	\end{pmatrix}
	\begin{pmatrix}
	\Phi \\ \sigma
	\end{pmatrix}.
\end{align}
One can see that, while this transformation does not alter the kinetic term of Eq.~\eqref{eq:lsm-general}, the potential \emph{does} change, implying redundancies in the parameters.

The rest of this section is devoted to show that any set of parameters satisfying Eq.~\eqref{eq:prm-inf} is equivalent to $a = c = 1/2$ and $b = (6 \xi + 1)/2$ because of this redundancy related to the O$(1,1)$ transformation.
The second term in the potential \eqref{eq:potential} transforms as follows:
\begin{align}
	&\frac{1}{144 \alpha} \left\{ \left( \Phi + \sigma \right) \left[ (2a - 1) \Phi - \sigma  \right] - b \phi_{i}^{2} \right\}^{2} \nonumber \\
	= &
	\frac{1}{144 \alpha} \left\{ \left( \Phi' + \sigma' \right) \left[ \left(a + (a - 1) e^{2\theta} \right) \Phi' - \left(a - (a - 1) e^{2\theta} \right)\sigma'  \right] - b \phi_{i}^{2} \right\}^{2}.
\end{align}
We can take a particular $\theta_{0}$ such that $a = (1 - a) e^{2 \theta_{0}}$ for $0 < a < 1$ which is automatic in Eq.~\eqref{eq:prm-inf}. 
Then the second term of the potential becomes
\begin{align}
	\frac{1}{144 \alpha} \left[ 2 a \sigma' \left( \Phi' + \sigma' \right) - b \phi_{i}^{2} \right]^2.
\end{align}
One can see that the potential takes exactly the same form as $a = c = 1/2$ and $b = (6 \xi + 1)/2$ after the following redefinition:
\begin{align}
	b \to  a (6 \xi + 1) , \quad \alpha \to  4 a^{2} \alpha .
\end{align}
Now it is clear that the potential has a minimum at $\sigma' = \phi_{i} = 0$ while it asymptotically approaches a constant value for $\sigma', \Phi' \to \infty$ and $\phi_{i}^{2} \to \text{const.}$ under $\Phi^{'2} - \sigma^{'2} - \phi_{i}^{2} = 6 M_{P}^{2}$, which is suitable for inflation.

\small
\bibliographystyle{utphys}
\bibliography{ref}

\end{document}